\documentclass[aps,floats,twocolumn,showpacs,prb,floatfix]{revtex4}
\usepackage{times}
\usepackage{graphicx,amsmath}
\newcommand{\mb}[1]{{\mathbf{#1}}}
\newcommand{\ek}{\epsilon_{\mathbf{k}}}

\newcommand{\sumk}{\sum_{\mathbf{k}}}
\newcommand{\phik}{\varphi_{\mathbf{k}}}
\newcommand{\phikq}{\varphi_{{\mathbf{k}}-{\mathbf{q}}/2}}
\newcommand{\dbar}[1]{\bar{\bar{#1}}}

\begin{document}
\title{Pairing fluctuation theory of high $T_c$ superconductivity in the
  presence of nonmagnetic impurities }

\author{Qijin Chen and J. R. Schrieffer} 

\affiliation{National High Magnetic Field Laboratory, Florida State
  University, Tallahassee, Florida 32310}

\date{\today}

\begin{abstract}
  The pseudogap phenomena in the cuprate superconductors requires a
  theory beyond the mean field BCS level. A natural candidate is to
  include strong pairing fluctuations, and treat the two-particle and
  single particle Green's functions self-consistently.  At the same
  time, impurities are present in even the cleanest samples of the
  cuprates. Some impurity effects can help reveal whether the pseudogap
  has a superconducting origin and thus test various theories.  Here we
  extend the pairing fluctuation theory for a clean system [Phys.  Rev.
  Lett. 81, 4708 (1998)] to the case with nonmagnetic impurities.  Both
  the pairing and the impurity $T$ matrices are included and treated
  self-consistently. We obtain a set of three equations for the chemical
  potential $\mu$, $T_c$, the excitation gap $\Delta(T_c)$ at $T_c$, or
  $\mu$, the order parameter $\Delta_{sc}$, and the pseudogap
  $\Delta_{pg}$ at temperature $T<T_c$, and study the effects of
  impurity scattering on the density of states, $T_c$ and the order
  parameter, and the pseudogap.  Both $T_c$ and the order parameter as
  well as the total excitation gap are suppressed, whereas the pseudogap
  is not for given $T\le T_c$.  Born scatterers are about twice as
  effective as unitary scatterers in suppressing $T_c$ and the gap.  In
  the strong pseudogap regime, pair excitations contribute a new
  $T^{3/2}$ term to the low $T$ superfluid density. The initial rapid
  drop of the zero $T$ superfluid density in the unitary limit as a
  function of impurity concentration $n_i$ also agrees with experiment.
\end{abstract}
\pacs{
74.20.-z, 
74.25.Fy, 
74.25.-q 
\hfill \textsf{cond-mat/0202541}
}
\maketitle

\section{Introduction}

The pseudogap phenomena in high $T_c$ superconductors have been a great
challenge to condensed matter physicists since over a decade ago. These
phenomena manifestly contradict BCS theory by, e.g, presenting a pseudo
excitation gap in single particle excitation spectrum. Yet the origin of
the pseudogap and, in general, the mechanism of the superconductivity
are still not clear. Many theories have been proposed, which fall into
two classes, based on whether the pseudogap has a superconducting
origin. Some authors propose that the pseudogap may not be related to
the superconductivity; instead, it is associated with another ordered
state, such as the antiferromagnetism related resonating valence bond
(RVB) state,\cite{Anderson} d-density wave\cite{Laughlin} and spin
density wave order.\cite{Sachdev} On the other hand, many others believe
that the pseudogap has the same origin as the superconductivity, such as
the phase fluctuation scenario of Emery and Kivelson\cite{Emery} and the
various precursor superconductivity
scenarios.\cite{Randeria,Maly1,Maly2,Janko,Chen-PRL98,Kosztin-PRB98}
Previously, Chen and coworkers have worked out, within the precursor
conductivity school, a pairing fluctuation
theory\cite{Chen-PRL98,Chen-PRB99,Chen-PRL00} which enables one to
calculate quantitatively physical quantities such as the phase diagram,
the superfluid density, etc. for a clean system. In this theory,
two-particle and one-particle Green's functions are treated on an equal
footing, and equations are solved self-consistently. Finite
center-of-mass momentum pair excitations become important as the pairing
interaction becomes strong, and lead to a pseudogap in the excitation
spectrum. In this context, these authors have been able to obtain a
phase diagram and calculate the superfluid density, in
(semi)quantitative agreement with experiment.

However, to fully apply this theory to the cuprates, we need to extend
it to impurity cases, since impurities are present even in the cleanest
samples of the high $T_c$ materials, such as the optimally doped
YBa$_2$Cu$_3$O$_{7-\delta}$ (YBCO) single crystals. In addition, this is
necessary in order to understand the finite frequency conductivity
issue. Furthermore, study of how various physical quantities respond to
impurity scattering may help to reveal the underlying mechanism of the
superconductivity. For example, it can be used to determine whether the
pseudogap has a superconducting origin.\cite{Balatsky} Particularly, it
is important to address how $T_c$ and the pseudogap itself vary with
impurity scattering, especially in the underdoped regime. To this end,
one needs to go beyond BCS theory and include the pseudogap as an
intrinsic part of the theory. Due to the complexity and technical
difficulties of this problem, there has been virtually no work in the
field on this important problem.

Among all physical quantities, the density of states (DOS) $N(\omega)$
close to the Fermi level ($\omega=0$) is probably most sensitive to the
impurities.  Yet different authors have yielded contradictory results in
this regard.  BCS-based impurity $T$ matrix calculations predict a
finite DOS at $\omega=0$,\cite{Hirschfeld86,Hirschfeld88} which has been
used to explain the crossover from $T$ to $T^2$ power law for the low
temperature superfluid density.\cite{Goldenfeld} Nonperturbative
approaches have also been studied and have yielded different results.
Senthil and Fisher\cite{Fisher} find that DOS vanishes according to
universal power laws, P\'epin and Lee\cite{Lee} predict that $N(\omega)$
diverges as $\omega \rightarrow 0$, assuming a strict particle-hole
symmetry, and Ziegler\cite{Ziegler} and coworkers' calculation shows a
rigorous lower bound on $N(\omega)$. Recently, Atkinson \textit{et
  al.}\cite{Atkinson} try to resolve these contradictions by fine-tuning
the details of the disorders.  Nevertheless, all these calculations are
based on BCS theory and cannot include the pseudogap in a
self-consistent fashion, and thus can only be applied to the low $T$
limit in the underdoped cuprates.  Therefore, it is necessary to extend
the BCS-based calculations on impurity issues to include the pseudogap
self-consistently.

In this paper, we extend the pairing fluctuation theory from clean to
impurity cases. Both the impurity scattering and particle-particle
scattering $T$ matrices are incorporated and treated self-consistently.
This goes far beyond the usual self-consistent (impurity) $T$-matrix
calculations at the BCS level by, e.g., Hirschfeld and
others.\cite{Hirschfeld86,Hirschfeld88} In this context, we study the
evolution of $T_c$ and various gap parameters as a function of the
coupling strength, the impurity concentration, the hole doping
concentration and the impurity scattering strength.  In addition, we
study not only the Born and the unitary limits, but also at intermediate
scattering strength. We find that the real part of the frequency
renormalization can never be set to zero, the chemical potential adjusts
itself with the impurity level. As a consequence, the positive and
negative strong scattering limits do not meet. The residue density of
states at the Fermi level is generally finite at finite impurity
concentrations, in agreement with what has been observed experimentally.
Both $T_c$ and the total excitation gap decrease with increasing
impurity level, as one may naively expect.  Born scatterers are about
twice as effective as unitary scatters in suppressing $T_c$ and the gap.
In the unitary limit, the zero temperature superfluid density decreases
faster with $n_i$ when $n_i$ is still small, whereas in the Born limit,
it is the opposite. At given $T<T_c$, both the order parameter and the
total gap are suppressed, but the pseudogap is not. Finally, incoherent
pair excitations contribute an additional $T^{3/2}$ term to the low $T$
temperature dependence of the superfluid density, robust against
impurity scattering.

In the next section, we first review the theory in a clean system, and
then present a theory at the Abrikosov-Gor'kov level. Finally, we
generalize it to include the full impurity $T$-matrix, in addition to
the particle-particle scattering $T$-matrix, in the treatment, and
obtain a set of three equations to solve for $\mu$, $T_c$ and various
gaps.  In Sec.~\ref{sec:result}, we present numerical solutions to
these equations. We first study the effects of impurity scattering on
the density of states, then study the effects on $T_c$ and the pseudogap
at $T_c$, followed by calculations of the effects on the gaps and the
superfluid density below $T_c$. Finally, we discuss some related issues,
and conclude our paper.

\section{Theoretical Formalism}
\label{sec:theory}

The excitation gap forms as a consequence of Cooper pairing in BCS
theory, while the superconductivity requires the formation of the
zero-momentum Cooper pair condensate. As these two occurs at the same
temperature in BCS theory, one natural way to extend BCS theory is to
allow pair formation at a higher temperature ($T^*$) and the Bose
condensation of the pairs at a lower temperature ($T_c$). Therefore,
these pairs are phase incoherent at $T>T_c$, leading to a pseudogap
without superconductivity. This can nicely explain the existence of the
pseudogap in the cuprate superconductors. Precursor superconductivity
scenarios, e.g., the present theory, provides a natural extension of
this kind. At weak coupling, the contribution of incoherent pairs is
negligible and one thus recovers BCS theory, with $T^*=T_c$. As the
coupling strength increases, incoherent pair excitations become
progressively more important, and $T^*$ can be much higher than $T_c$,
as found in the underdoped cuprates. In general, both fermionic
Bogoliubov quasiparticles and bosonic pair excitations coexist at finite
$T<T_c$.

\subsection{Review of the theory in a clean system}
\label{subsec:review}

The cuprates can be modeled as a system of fermions which have an
anisotropic lattice dispersion $\epsilon_{\mathbf{k}}=2t_\parallel
(2-\cos k_x -\cos k_y) + 2t_\perp (1-\cos k_z) -\mu $, with an
effective, short range pairing interaction $V_{\mathbf{k,k'}} =
g\varphi_{\mathbf{k}} \varphi_{\mathbf{k'}}$, where $g<0$. Here
$t_\parallel$ and $t_\perp$ are the in-plane and out-of-plane hopping
integrals, respectively, and $\mu$ is the fermionic chemical potential.
For the cuprates, $t_\perp \ll t_\parallel$. The Hamiltonian is given by
\begin{eqnarray}
{\cal H}^0 & = & \sum_{{\bf k}\sigma} \epsilon_{\bf k}
c^{\dag}_{{\bf k}\sigma} c^{\ }_{{\bf k}\sigma}
\nonumber \\
& & + \sum_{\bf k k' q} V_{\bf k, k'} 
c^{\dag}_{{\bf k}+{\bf q}/2\uparrow} 
c^{\dag}_{-{\bf k}+{\bf q}/2\downarrow} 
c^{\ }_{-{\bf k'}+{\bf q}/2\downarrow} 
c^{\ }_{{\bf k'}+{\bf q}/2\uparrow},
\label{Hamiltonian}
\end{eqnarray}
The pairing symmetry is given by $\varphi_{\mathbf{k}}=1$ and $(\cos k_x
-\cos k_y)$ for $s$- and $d$-wave, respectively. Here and in what
follows, we use the superscript ``0'' for quantities in the clean
system, to be consistent with the notations for the impurity dressed
counterpart below. For brevity, we use a four-momentum notation:
$K=({\bf k}, i\omega), \sum_K = T \sum _ {{\bf k}, \omega}$, etc.

\begin{figure}
\includegraphics[width=3.3in]{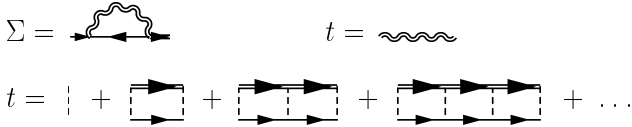}
\caption{Diagrams for the Dyson's equations in a clean system.}
\label{dyson.eq.clean}
\end{figure}

To focus on the superconductivity, we consider only the pairing channel,
following early work by Kadanoff and Martin;\cite{Kadanoff} the
self-energy is given by multiple particle-particle scattering. The
infinite series of the equations of motion are truncated at the
three-particle level $G_3$, and $G_3$ is then factorized into single-
($G$) and two-particle ($G_2$) Green's functions. The final result is
given by the Dyson's equations for the single particle propagator [Refer
to Refs.~(\onlinecite{Chen-PRL98,Chen-PRL00}) for details]
\begin{subequations}
\label{eq:t3}
\begin{eqnarray}
  \label{eq:t3a}
  \Sigma^0(K) &=& {G^0_0}^{-1}(K)-{G^0}^{-1}(K) \nonumber\\
  &=& \sum_Q t^0(Q)\,G^0_0(Q-K)\,\varphi^2_{{\bf k}-{\bf q}/2} \;,
\end{eqnarray}
and the $T$ matrix (or pair propagator) 
\begin{equation}
  \label{eq:t3b}
  t^0(Q) \;=\;t^0_{sc}(Q)+t^0_{pg}(Q)\;, 
\end{equation}
with
\begin{equation}
t^0_{sc}(Q)=
  -\frac{\Delta_{sc}^2}{T}\delta(Q)\;,
\end{equation}
where $\Delta_{sc} \equiv 0$ at $T\ge T_c$,
and
\begin{equation}
t^0_{pg}(Q)=\frac{g}{1+g\,\chi^0(Q)}\;,
\end{equation}
where $\Delta_{sc}$ is the superconducting order parameter, $G^0_0(K) =
1/(i\omega-\epsilon_{\bf k})$ is the bare propagator. and
\begin{equation}
  \label{eq:t3c}
  \chi^0(Q) \;=\; \sum_K G^0(K)\,G^0_0(Q-K)\,\varphi^2_{{\bf k}-{\bf q}/2}
\end{equation}
\end{subequations}
is the pair susceptibility.  This result can be represented
diagrammatically by Fig.~\ref{dyson.eq.clean}. The single and double
lines denote the bare and full Green's functions, respectively, and the
wiggly double lines denote the pair propagator. 

The superconducting instability is given by the Thouless criterion
\begin{equation}
  \label{eq:thouless}
  1+g\chi^0(0)=0\;, \qquad (T\le T_c),
\end{equation}
which leads to the approximation
\begin{eqnarray}
  \label{eq:pg}
  \Sigma^0_{pg}(K) &=& \sum_Q
  t^0_{pg}(Q)\,G^0_0(Q-K)\,\varphi^2_{{\bf k}-{\bf q}/2} \\\nonumber 
  &\approx & -\Delta_{pg}^2  G^0_0(-K)\,\varphi^2_{\bf k} \;,
\end{eqnarray}
where the pseudogap is defined by
\begin{equation}
\Delta_{pg}^2 = - \sum_{Q\ne 0} t^0_{pg}(Q) \;.
\label{eq:PG-def}
\end{equation}
As a consequence, the self-energy takes the standard BCS form
\begin{equation}
\Sigma(K) = -\Delta^2 G_0^0 (-K)\varphi^2_{\bf k} = -\Delta^2_{\bf k}
G_0^0 (-K);
\end{equation}
where $\Delta^2 = \Delta_{sc}^2 + \Delta_{pg}^2$, and $\Delta_{\bf k}
= \Delta \varphi_{\bf k}$. In this way, the full
Green's function $G^0(K)$ also takes the standard BCS form, with the
quasiparticle dispersion given by $E_{\bf k}= \sqrt{\epsilon_k^2+
  \Delta^2\varphi_{\bf k}^2}$. So does the excitation gap equation
\begin{equation}
  1+g\sum_{\mathbf{k}} \frac{1- 2f(E_{\mathbf{k}})}{2E_{\mathbf{k}}}\,
  \varphi^2_{\mathbf{k}} = 0 \:. 
\label{eq:gap-clean}
\end{equation}
We emphasize that although this equation is formally identical to its
BCS counterpart, the $\Delta$ here can no longer be interpretted as the
order parameter as $\Delta_{pg}\ne 0 $ in general.
For self-consistency, we have the fermion number constraint
\begin{equation}
  \label{eq:n}
  n = 2\sum_K G^0(K)= 
2\sum_{\mathbf{k}}\left[v_{\bf k}^2 +
   \frac{\epsilon_{\mathbf{k}}}{E_{\mathbf{k}}} f(E_{\mathbf{k}})\right]\:,
\end{equation}

The gap equation Eq.~(\ref{eq:gap-clean}), the number equation
Eq.~(\ref{eq:n}), and the pseudogap parametrization
Eq.~(\ref{eq:PG-def}) form a complete set, and can be used to solve
self-consistently for $T_c$, $\mu(T_c)$, and $\Delta(T_c)$ by setting
$\Delta_{sc}=0$, or $\mu(T)$, $\Delta(T)$, and $\Delta_{pg}(T)$ at given
$T<T_c$. Here $v_{\bf k}^2 = \frac{1}{2}(1-
\epsilon_{\mathbf{k}}/E_{\mathbf{k}})$, as in BCS.

\subsection{Impurity scattering at the Abrikosov-Gor'kov level}
\label{subsec:AG}

For simplicity, we restrict ourselves to nonmagnetic, elastic, isotropic
$s$-wave scattering. At the same time, we will keep the derivation as
general as possible.  In the presence of impurities of concentration
$n_i$, the Hamiltonian is given by
\begin{equation}
{\cal H}= {\cal H}^0 + {\cal H}_I\:,
\end{equation}
where in the real space the impurity term is given by
\begin{equation}
{\cal H}_I = \sum_i\int d\textbf{x}\; u(\textbf{x}-\textbf{x}_i)
\psi^\dagger(\textbf{x})\psi(\textbf{x}) \:,
\end{equation}
with $u(\textbf{x})=u\delta(\textbf{x})$ for isotropic $s$-wave scattering.

\begin{figure}
\centerline{\includegraphics[width=3.3in]{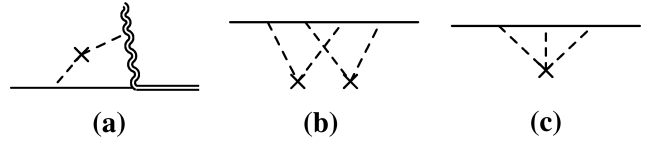}}
\caption{Examples of (a) bridging, (b) crossing diagrams, and 
(c) higher order terms neglected in the calculation.}
\label{diagram-excluded}
\end{figure}

\begin{figure*}
\centerline{\hspace*{1cm} \includegraphics[width=5.5in]{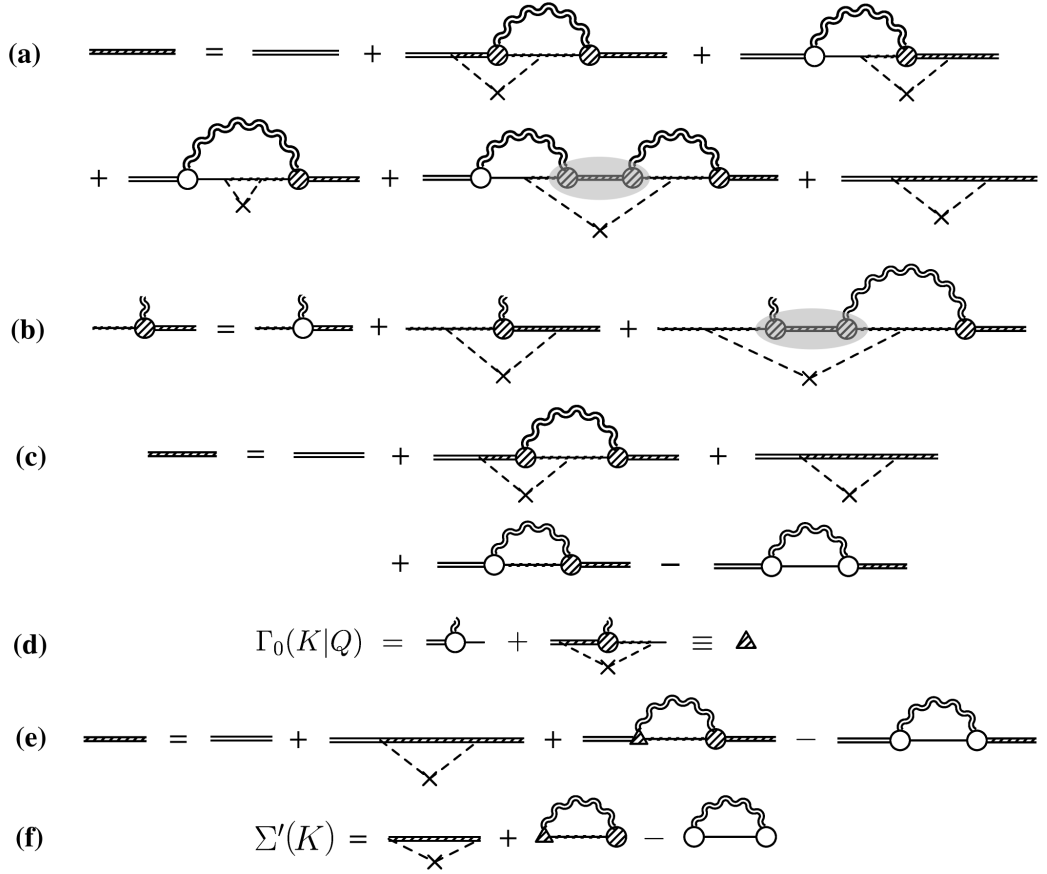}}
\caption{Feynman diagrams for the impurity dressed full Green's function.}
\label{diag-G1}
\end{figure*}

\begin{figure*}
\centerline{\includegraphics[width=5.5in]{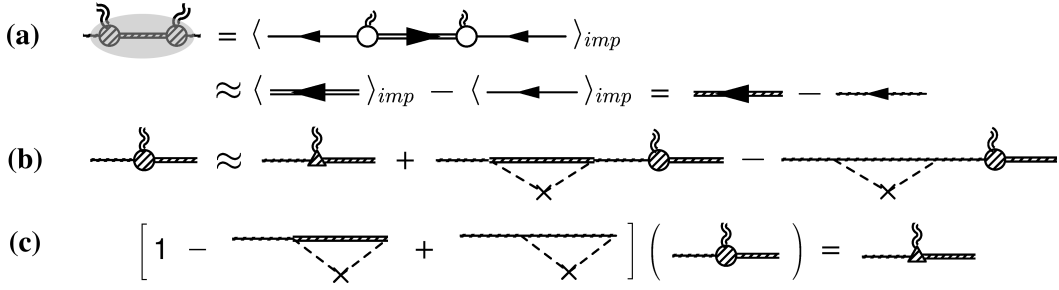}}
\caption{Feynman diagrams for pairing vertex.}
\label{diag-vertex}
\end{figure*}

To address the impurity scattering, we begin at the Abrikosov-Gor'kov
(AG) level,\cite{AG,AG2,AGD} which is a good approximation in the Born
limit.  Following AG, we include all possible configurations of impurity
dressing, but excluding bridging diagrams like
Fig.~\ref{diagram-excluded}(a), crossing diagrams like
Fig.~\ref{diagram-excluded}(b), and higher order terms like
Fig.~\ref{diagram-excluded}(c). The dashed lines denote impurity
scattering, and the crosses denote the impurity vertices. We clarity, in
most diagrams, we do not draw the fermion propagation arrows. It is
understood, however, that they change direction at and only at a pairing
vertex. As in subsec.~\ref{subsec:review}, we use plain double lines to
denote the Green's function ($G^0$) fully dressed by the pairing
interaction but without impurity scattering, i.e.,
\begin{equation}
G^0(K) = \frac{1}{i\omega - \epsilon_{\bf k} - \Sigma^0 (K) } \;,
\end{equation}
where
\begin{equation}
\Sigma^0(K) = \sum_Q t(Q) G^0_0 (Q-K) \phikq^2 \;.
\end{equation}
However, since we will address the impurity dressing of the pairing
vertex or, equivalently, the pair susceptibility $\chi(Q)$, we assume
the pair propagator $t(Q)$ in the above equation is already dressed with
impurity scattering, with
\begin{equation}
t(Q)=t_{sc}(Q)+t_{pg}(Q) \;,
\end{equation}
\begin{equation}
t_{sc}(Q)=  -\frac{\Delta_{sc}^2}{T}\delta(Q)\;,
\end{equation}
and

\begin{equation}
\Delta_{pg}^2 = -\sum_{Q\ne 0} t_{pg}(Q) = 
-\sum_{Q\ne 0} \frac{g}{1+g\chi(Q)} \:,
\label{eq:PG-imp}
\end{equation}
as in the clean case. The shaded double lines denote impurity-dressed
full Green's function $G$, and ``shaded'' single lines denote
impurity-dressed bare Green's function (which we call $\hat{G}_0$), i.e,
\begin{equation}
  \hat{G}_0(K) = \frac{1}{i\omega - \epsilon_{\bf k} -
    \bar{\hat{G}}_{0,\omega}}\;,
\end{equation}
where the bar denotes impurity average $\bar{\hat{G}}_{0,\omega} = n_i
\sum_{\bf k^\prime} |u(\textbf{k} -\textbf{k}^\prime)|^2
\hat{G}_0(K^\prime) $.  We use open circles to denote bare pairing
vertex $\gamma(K|Q)=\varphi_{\mb{k}+\mb{q}/2}$, and shaded circles full
pairing vertex $\Gamma(K|Q)$, where $Q$ is the pair four-momentum.  To
obtain the Feynman diagrams for the impurity dressed full $G$, we first
expand the pairing self-energy diagram as an infinite series which
contains only bare single particle Green's function and pair
propagators, and then insert all possible impurity scattering on the
single particle propagators at the AG level. We assume that the pair
propagators are always self-consistently dressed by the impurity
scattering. After regrouping all non-impurity dressed lines on the left,
the final result for the diagrams is shown in Fig.~\ref{diag-G1}(a).
Here following AG, the subdiagrams inside the two impurity legs are
assumed to be self-consistently dressed by impurity scattering.  To make
direct comparison with the BCS case easier, we present the corresponding
diagrams for the BCS case in Appendix \ref{App:BCS}. The first term on
the right hand side (RHS) of Fig.~\ref{diag-G1}(a) contains all diagrams
without impurity dressing (except via the pair propagators). The second
term corresponds to the third term of the first equation in
Fig.~\ref{diag:BCS-AG}. The third term corresponds to the last term, the
last term to the second. The fourth and the fifth together correspond to
the fourth term [see Fig.~\ref{diag-vertex}(a)].  The fifth term in
Fig.~\ref{diag-G1}(a) arises since the two impurity legs can cross two
separate pairing self-energy dressing parts; it can be eliminated using
the equality shown in Fig.~\ref{diag-G1}(b). Here the shaded elliptical
region denotes self-consistent impurity dressing of the double pairing
vertex structure inside the two impurity scattering legs, as shown in
Fig.~\ref{diag-vertex}(a). It is worth pointing out that these diagrams
reduce to their BCS counterpart if one removes the pairing propagators.
The Dyson's equation $\hat{G}_0^{-1}(K) = {G_0^0}^{-1}(K) -
\bar{\hat{G}}_{0,\omega}$ can then be used to eliminate the fourth term
in Fig.~\ref{diag-G1}(a).  We now obtain the greatly simplified diagrams
for $G$ as shown in Fig.~\ref{diag-G1}(c), which can be further reduced
into Fig.~\ref{diag-G1}(e), upon defining a reduced pairing vertex
$\Gamma_0(K|Q)$ (shaded triangles) as shown in Fig.~\ref{diag-G1}(d),

\begin{eqnarray}
 \hspace*{-5ex} \Gamma_0(K|Q) &=& \phikq \nonumber\\
 &&\hskip -1.6cm + n_i \sum_{\bf k^\prime} |u(\textbf{k}
  -\textbf{k}^\prime)|^2 \hat{G}_0 (Q-K^\prime) \Gamma (K^\prime|Q)
  G(K^\prime) \;.
\end{eqnarray}
One can then read off the impurity-induced \textit{quasiparticle}
self-energy $\Sigma^\prime (K) = {G^0}^{-1}(K) - G^{-1}(K)$ immediately,
as shown in Fig.~\ref{diag-G1}(e),
\begin{equation}
\Sigma^\prime(K)  = \bar{G}_\omega +
\Sigma(K) - \Sigma_0(K) \;,
\end{equation}
where the impurity average
\begin{equation}
\bar{G}_\omega = n_i \sum_{\bf k^\prime}
|u(\textbf{k} -\textbf{k}^\prime)|^2 G(K^\prime) \;, 
\label{eq:Gwbar}
\end{equation}
and the ``full'' self-energy
\begin{equation}
\Sigma(K) = \sum_Q \Gamma_0(K|Q) t(Q) \Gamma(K|Q) \hat{G}_0 (Q-K) \;.
\end{equation}
Therefore, we have finally
\begin{eqnarray}
G^{-1}(K) &=& {G^0}^{-1}(K) -\Sigma^\prime (K) = i\omega - \ek -
\bar{G}_\omega -\Sigma(K) \nonumber\\
&=& i\tilde\omega - \ek -\Sigma (K) = G_0^{-1}(K) - \Sigma(K) \;,
\end{eqnarray}
where we have defined the renormalized frequency $i\tilde\omega = i\omega -
\bar{G}_\omega$ and the ``bare'' Green's function 
\begin{equation}
G_0(K) = \frac{1}{i\tilde\omega -\ek} \;.
\end{equation}
Note here $G_0(K) \ne \hat{G}_0(K)$.

Now we deal further with the pairing vertex $\Gamma(K|Q)$. First, we
notice that the impurity-dressed double-vertex structure in
Fig.~\ref{diag-G1}(a) can be simplified as shown in
Fig.~\ref{diag-vertex}(a) using the approximation
\begin{equation}
\sum_Q t(Q) f(K, Q) \approx \bigg[\sum_Q t(Q)\bigg] f(K, 0) \;,
\label{eq:approx}
\end{equation}
where $f(K,Q)$ is an arbitrary slow-varying function of $Q$. This is due
to the fact that $t(Q)$ diverges as $Q \rightarrow 0$ at $T\le T_c$.
The Dyson's equation for the Green's function $G^0(-K)$ in a clean
system is also used in getting the second line of
Fig.~\ref{diag-vertex}(a). Therefore, we have approximately the
impurity-dressed pairing vertex as shown in Fig.~\ref{diag-vertex}(b),
which implies the equality shown in Fig.~\ref{diag-vertex}(c).
Fig.~\ref{diag-vertex}(c) can be written as
\begin{eqnarray}
  \hat{G}_0(K) \Gamma(K|Q) G(Q-K) &=&
  \frac{\Gamma_0(K|Q)G(Q-K)}{\hat{G}_0^{-1}(K) - \bar{G}_\omega +
    \bar{\hat{G}}_{0,\omega}} \nonumber\\ 
& & \hskip -0.7in = G_0(K)
  \Gamma_0(K|Q)G(Q-K) \;.
\end{eqnarray}
This result demonstrates the following important relationship:
\begin{equation}
\hat{G}_0(K) \Gamma(K|Q) = G_0(K)\Gamma_0(K|Q), \quad (\mbox{small\ } Q) \;.
\label{eq:vertex-relation}
\end{equation}
Using this relationship, now the self-energy can be simplified as
follows:
\begin{eqnarray}
  \Sigma(K) &=& \sum_Q \Gamma_0^2(K|Q) t(Q) G_0(Q-K) \nonumber\\ 
  &\approx& -\Delta^2 \Gamma_0^2(K) G_0(-K) = -\tilde{\Delta}_{\bf k}^2
  G_0(-K) \;,
\end{eqnarray}
where $\Gamma_0(K) \equiv \Gamma(K|Q=0)$ and $\tilde{\Delta}_{\bf k} = \Delta
\Gamma_0(K)$.  

Finally, the impurity dressing of each rung [i.e., $\chi(Q)$] of the
particle-particle scattering ladder diagrams is topologically identical
to the impurity dressing of the pairing vertex and the two associated
single particle lines. And summing up all the ladders gives the pairing
$T$ matrix. Therefore, the pair susceptibility becomes
\begin{eqnarray}
\chi(Q) &=& \sum_K \Gamma(K|Q) \hat{G}_0(Q-K) G(K) \phikq \nonumber\\
&& \approx \sum_K \Gamma_0(K|Q) G_0(Q-K) G(K) \phikq \;.
\end{eqnarray}
And the gap equation is given by
\begin{equation}
1+g\chi(0)=0 = 1+g\sum_K \Gamma_0(K) G(K)G_0(-K) \phik \;.
\label{eq:gap-imp}
\end{equation}
This result can be easily verified to be consistent with the
self-consistency condition. Define formally the generalized Gor'kov $F$
function:
\begin{equation}
F^\dagger(K) \equiv \Delta \hat{G}_0(-K) \Gamma(K) G(K) \;.
\end{equation}
Using Eq.~(\ref{eq:vertex-relation}), we have 
\begin{equation}
  F^\dagger(K) \equiv \Delta\Gamma_0(K) G_0(-K) G(K) =
  \tilde{\Delta}_{\bf k} G_0(-K) G(K)\;.
\end{equation}
%
The \textit{formal} difference between this $F^\dagger$ and that in BCS
is that $\phik$ is now replaced by the renormalized vertex
$\Gamma_0(K)$.  One immediately sees that the condition
\begin{equation}
  \Delta_{\bf k} 
\equiv -g \sum_{K^\prime} \phik
  \varphi_{\bf k^\prime} F^\dagger (K^\prime)
\end{equation}
is consistent with the gap equation Eq.~(\ref{eq:gap-imp}). However, it
should be emphasized that the $F$ function so defined does not vanish
above $T_c$ in the pseudogap regime, different from the BCS case.


\subsection{Impurity scattering beyond the AG level}

\begin{figure}
\centerline{\includegraphics[width=3in]{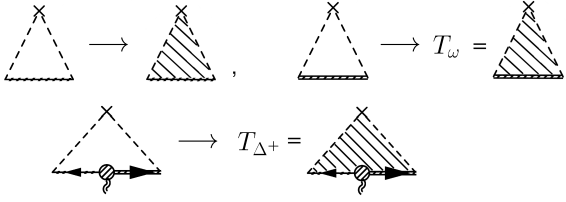}}
\caption{Replacement scheme from the AG level to self-consistent
  impurity $T$-matrix treatment.}
\label{diag-replacement}
\end{figure}

In this subsection, we include both the impurity scattering $T$-matrix
with the particle-particle scattering $T$-matrix, and, thus, go beyond
the AG level. We notice that if one replaces the second-order impurity
scattering subdiagrams at the AG level with the corresponding impurity
$T$-matrices, as shown in Fig.~\ref{diag-replacement}, the derivation
for $G(K)$ goes through formally without modification. Now we only need
to determine the impurity $T$-matrices $T_\omega$ and $T_{\Delta^\dag}$
(as well as their complex conjugate) in terms of their AG-level
counterpart, $\bar{G}_\omega$ and $\bar{F}^\dagger_\omega = n_i
\sum_{\bf k^\prime} |u({\bf k-k^\prime})|^2 F^\dagger(K^\prime) $,
respectively.
In other words, except that $i\tilde\omega$ and $\tilde{\Delta}_{\bf k}$
now have different expressions, everything else remains the same in terms of
$i\tilde\omega$ and $\tilde{\Delta}_{\bf k}$ (as well as their complex
conjugate), just as in the BCS case (see Appendix \ref{App:BCS}).

\begin{figure*}
\centerline{\includegraphics[width=4.7in]{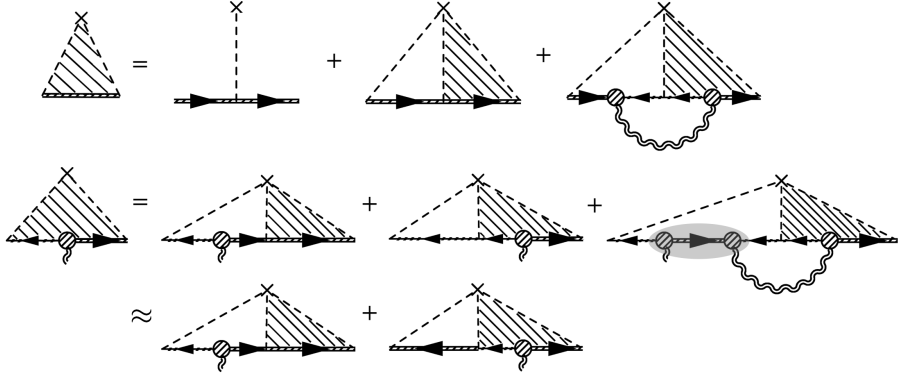}}
\caption{Relationship between the regular impurity $T$-matrix $T_\omega$
  and the anomalous impurity $T$-matrix $T_{\Delta^\dagger}$.}
\label{diag-Tmatrix}
\end{figure*}

The Feynman diagrams for $T_\omega$ and $T_{\Delta^\dag}$ are shown in
Fig.~\ref{diag-Tmatrix}. To obtain the second line for
$T_{\Delta^\dag}$, we make use of the approximation in
Fig.~\ref{diag-vertex}(a) to convert the left part of the second and the
third term on the first line to the full $G$.  This result is direct
analogy with its BCS counterpart as shown in
Fig.~\ref{diag:BCS-Tmatrix}.  One can now write down the equations for
$T_\omega$ and $T_{\Delta^\dag}$ without difficulty.
\begin{eqnarray}
T_\omega(\mb{k}, \mb{k}^\prime) &=& u (\mb{k}, \mb{k}^\prime) + \sum_{\bf
  k''} u (\mb{k}, \mb{k}'') G(K'') T_\omega(\mb{k}'', \mb{k}^\prime) 
\nonumber\\
&& + \sum_{\bf k''}\sum_Q u (\mb{k}, \mb{k}'') G(K'')
  \hat{G}_0(Q-K'') 
\nonumber\\
&&{}\times
\Gamma(K''|Q) t(Q) T_{\Delta^\dag}(K''-Q, K^\prime|Q) \;,
\end{eqnarray}
and
\begin{eqnarray}
\lefteqn{T_{\Delta^\dag}(K-Q, K^\prime|Q) = \sum_{\bf k''} u
  (\mb{k}''-\mb{q}, \mb{k}-\mb{q}) } \nonumber
\\ && {}\times
\hat{G}_0(Q-K'') G(K'')\Gamma(K''|Q) T_\omega(\mb{k}'', \mb{k}^\prime)
\nonumber\\&& + \sum_{\bf k''} u (\mb{k}''-\mb{q}, \mb{k}-\mb{q}) G(Q-K'')
T_{\Delta^\dag}(K''-Q, K^\prime|Q) \,. \nonumber\\
\end{eqnarray}
Note here $T_{\Delta^\dag}$ does not contain the factor $\Delta$, unlike
its BCS counterpart. It has the same dimension as $\Gamma$. Both
$\Gamma$ and $\Gamma_0$ now contain the full impurity $T$-matrix beyond
the AG level, and the vertex relation Eq.~(\ref{eq:vertex-relation})
remains valid. The new expression for $\Gamma_0$ is given by
\begin{equation}
\Gamma_0(K|Q) = \phikq + n_i \sum_{\mb{k}^\prime} T_{\Delta^\dag}(K-Q,
K^\prime|Q) \:.
\end{equation}

So far, we have kept the derivation for a generic elastic scattering
$u(\mb{k},\mb{k}^\prime)$. For isotropic $s$-wave scattering,
$u(\mb{k},\mb{k}^\prime)=u$. In this case, $T_\omega$ and
$T_{\Delta^\dag}$ are independent of $\mb{k}$ and
$\mb{k}^\prime$. Neglecting the momentum dependence, we obtain
\begin{subequations}
\begin{equation}
T_{\Delta^\dag}(\omega-\Omega, \omega) = \frac{u \sum_{\bf k}
  G_0(Q-K)\Gamma_0(K|Q) G(K)}{1-u\sum_{\bf k} G(Q-K)} T_\omega \;,
\end{equation}
and
\begin{eqnarray}
T_\omega &=& u + u \sum_{\mb{k}} G(K) T_\omega \nonumber \\
 &&{} + u \sum_Q t(Q) \bigg[  \sum_{\mb{k}} G(K) \Gamma_0(K|Q) G_0
 (Q-K) \bigg] \nonumber\\
&&{}\times T_{\Delta^\dag}(\omega-\Omega, \omega) \;.
\end{eqnarray}
\end{subequations}
Use has been made of the vertex relation Eq.~(\ref{eq:vertex-relation}).
Here again, we need to make use of the approximation Eq.~(\ref{eq:approx}).
Defining $\bar{\bar{G}}_\omega = \sum_{\mb{k}}G(K)$ and
\begin{equation}
\dbar{F}^\dagger_\omega = \sum_{\mb{k}} F^\dagger (K) = \sum_{\mb{k}} \Delta
\Gamma_0(K) G_0(-K)G(K) \,,
\label{eq:Fw}
\end{equation}
we obtain
\begin{subequations}
\begin{equation}
T_\omega = \frac{u (1-u\dbar{G}_{-\omega})}{\left( 1-u
    \dbar{G}_\omega\right) \left( 1-u
    \dbar{G}_{-\omega}\right) + u^2 \dbar{F}_\omega
    \dbar{F}^\dagger_\omega} \,, 
\label{eq:Tw}
\end{equation}
and
\begin{eqnarray}
T_{\Delta^\dag} (\omega-\Omega, \omega) &=& \frac{u^2 \sum_{\bf
    k} G_0(Q-K) \Gamma_0 (K|Q) G(K)} {\left( 1-u
    \dbar{G}_\omega\right) \left( 1-u
    \dbar{G}_{-\omega}\right) + u^2 \dbar{F}_\omega
    \dbar{F}^\dagger_\omega }\nonumber \\ &&{}\times \frac{ 1-u
    \dbar{G}_{-\omega} }{  1-u
    \dbar{G}_{\Omega-\omega} } \,.
\end{eqnarray}
Letting $\Omega\rightarrow 0$, the last equation becomes
\begin{equation}
\Delta T_{\Delta^\dag}(\omega) = \frac{u^2 \dbar{F}^\dagger_\omega} {\left( 1-u
    \dbar{G}_\omega\right) \left( 1-u
    \dbar{G}_{-\omega}\right) + u^2 \dbar{F}_\omega
    \dbar{F}^\dagger_\omega } \,.
\label{eq:TDelta}
\end{equation}
\end{subequations}

The frequency and gap renormalizations are given by
\begin{subequations}
\begin{equation}
i\tilde{\omega} = i\omega - \Sigma_\omega \,, \quad
i\tilde{\underline{\omega}} = -i\omega - \Sigma_{-\omega} \,,
\end{equation}
\begin{equation}
\tilde{\Delta}_\mb{k} = \Delta_\mb{k} + \Sigma_\Delta \,, \quad
\tilde{\Delta}_\mb{k}^* = \Delta_\mb{k}^* + \Sigma_{\Delta^\dag} 
\end{equation}
\end{subequations}
where $\Sigma_\omega = n_i T_\omega$ and $\Sigma_\Delta = n_i \Delta
T_\Delta$. Here $\tilde{\underline{\omega}}=  \widetilde{(-\omega)}$. 
The expression for $\chi(Q)$ remains the same
as in previous subsection.

For $d$-wave, $T_{\Delta^\dag}=T_\Delta = 0$, and $\tilde{\Delta}_\mb{k} =
\Delta_\mb{k}$. Then Eq.~(\ref{eq:Tw}) is greatly simplified,
\begin{equation}
T_\omega = \frac{u}{1-u \dbar{G}_\omega} \,.
\label{eq:Tfinal}
\end{equation}
The full Green's function is given by
\begin{equation}
G(K) = \frac{i\tilde{\underline{\omega}} -\ek}
{(i\tilde{\omega}-\ek) (i\tilde{\underline{\omega}} -\ek)  +
  \Delta_\mb{k}^* \Delta_\mb{k}} \,.
\label{eq:Gfinal}
\end{equation}

Due to the approximation Eq.~(\ref{eq:approx}), we are able to bring the
final result Eqs.~(\ref{eq:Tfinal}) and (\ref{eq:Gfinal}) into the BCS
form. It is easy to show that they are equivalent to the more familiar
form in Nambu formalism, as used in Ref.~\onlinecite{Hirschfeld88}.
Define
\begin{equation}
T_{A\omega} = \frac{1}{2} (T_\omega - T_{-\omega}), \quad T_{S\omega} =
\frac{1}{2} (T_\omega + T_{-\omega})\,, 
\end{equation}
and similarly for $\Sigma_\omega$ and $\dbar{G}_\omega$. Here the
subscript ``A'' and ``S'' denote antisymmetric and symmetric part,
respectively. Further define
\begin{equation}
i\tilde{\omega}_A = i\omega - \Sigma_{A\omega}, \quad
\tilde{\epsilon}_K = \ek + \Sigma_{S\omega} \,, 
\end{equation}
then we obtain (with $\Delta^* = \Delta$)
\begin{equation}
G(K) = \frac{i\tilde{\omega}_A + \tilde{\epsilon}_K}
{(i\tilde{\omega}_A)^2 - \tilde{\epsilon}_K^2 - \Delta_\mb{k}^2} \,,
\end{equation}
and
\begin{subequations}
\begin{eqnarray}
T_{A\omega} &=& \frac{u^2 \dbar{G}_{A\omega}} {(1-u\dbar{G}_{S\omega})^2 -
  u^2  \dbar{G}_{A\omega}^2 }\;, \\
T_{S\omega} &=& \frac{u (1-\dbar{G}_{S\omega})} {(1-u\dbar{G}_{S\omega})^2 -
  u^2  \dbar{G}_{A\omega}^2 }\;, 
\end{eqnarray}
\end{subequations}
where 
\begin{subequations}
\begin{eqnarray}
\dbar{G}_{A\omega} &=& \sumk \frac{i\tilde{\omega}_A }
{(i\tilde{\omega}_A)^2 - \tilde{\epsilon}_K^2 - \Delta_\mb{k}^2}
\,,\\ 
\dbar{G}_{S\omega} &=& \sumk \frac{\tilde{\epsilon}_K}
{(i\tilde{\omega}_A)^2 - \tilde{\epsilon}_K^2 - \Delta_\mb{k}^2} \,,
\end{eqnarray}
\end{subequations}
are the antisymmetric and symmetric parts of $\dbar{G}_\omega$,
respectively. It is evident that $T_A$ and $T_S$ correspond to $T_0$ and
$-T_3$, respectively, in the Nambu formalism in
Ref.~\onlinecite{Hirschfeld88}, (and similarly for $G_A$ and $G_S$).

It should be emphasized, however, that unless $u=0$ or $u=\pm \infty$,
the symmetric part of the impurity $T$-matrix, $T_{S\omega}$, can never
be set to zero, even if one could in principle have $\dbar{G}_{S\omega}
= 0$. This means that $\tilde{\epsilon}_K$ will always acquire a
non-trivial, frequency dependent renormalization, $\Sigma_{S\omega}$.
While this renormalization is small for weak coupling BCS
superconductors, it is expected to be significant for the cuprate
superconductors.

\section{Numerical solutions for $d$-wave superconductors}
\label{sec:result}

\subsection{Analytical continuation and equations to solve}
\label{subsec:continuation}

Since there is no explicit pairing vertex renormalization for $d$-wave
superconductors, i.e., $\Gamma_0(K) = \phik$ or
$\tilde{\Delta}_\mb{k}=\Delta_\mb{k}$, a major part of the numerics is
to calculate the frequency renormalization. Everything else will follow
straightforwardly.

Numerical calculations can be done in the real frequencies, after proper
analytical continuation.  Since $T_{S\omega} \ne 0$, $\Sigma_\omega$ and
$\Sigma_{-\omega}$ are independent of each other. To obtain the
frequency renormalization $\Sigma_\omega$, one has to solve a set of
four equations for $\dbar{G}_{\omega}$, $\dbar{G}_{-\omega}$,
$\Sigma_{\omega}$, and $\Sigma_{-\omega}$ self-consistently for given
$\omega$. Because $i\tilde{\underline{\omega}} \ne - i \tilde{\omega}$,
and one needs to analytically continue both simultaneously, the
analytical continuation must be done carefully.
For $n> 0$, $i\tilde{\omega}_n \rightarrow 
\omega_+^R = \omega_+ + i\Sigma''_+ $, and
$i\tilde{\underline{\omega}}_n \rightarrow 
\omega_-^A = \omega_- - i \Sigma''_- $.  For $n^\prime = -n < 0$,
$i\tilde{\omega}_{n^\prime} \rightarrow 
\omega_-^R = \omega_- + i \Sigma''_-$ and
$i\tilde{\underline{\omega}}_{n^\prime} \rightarrow 
\omega_+^A = \omega_+ - i\Sigma''_+$. Here $\omega_\pm = \pm \omega -
\Sigma^\prime_\pm$, and we choose $\omega> 0$ and $\Sigma_\pm''>0$.
Then we obtain four equations as follows:
\begin{eqnarray}
\dbar{G}^R_{\omega> 0} &=& \sumk \frac{\omega_- - i\Sigma_-'' -\ek}
{(\omega_+ + i \Sigma_+'' -\ek) (\omega_- - i\Sigma_-'' -\ek) +
  \Delta^2_\mb{k}} \nonumber\\
\dbar{G}^R_{-\omega< 0} &=& \sumk \frac{\omega_+ - i\Sigma_+'' -\ek}
{(\omega_- + i \Sigma_-'' -\ek) (\omega_+ - i\Sigma_+'' -\ek) +
  \Delta^2_\mb{k}} \nonumber\\
\Sigma^R_{\omega>0} &=& \frac{n_i u} {1 - u \dbar{G}^R_{\omega}} =
\Sigma_+^\prime - i \Sigma_+'' \nonumber\\
\Sigma^R_{-\omega>0} &=& \frac{n_i u} {1 - u \dbar{G}^R_{-\omega}} =
\Sigma_-^\prime - i \Sigma_-'' 
\label{eq:renorm}
\end{eqnarray}
These equations are solved self-consistently for $(\Sigma^\prime_\pm,
\Sigma''_\pm)$, as well as the real and imaginary parts of
$\dbar{G}_{\pm \omega}$, as a function of $\omega$. No Kramers-Kronig
relations are invoked in these numerical calculations. Note in real
numerics, we subtract $\Sigma^\prime_{\omega=0}$ from $\Sigma^R_\omega$
so that $\Sigma^\prime_{\omega=0}=0$. This subtraction is compensated by a
constant shift in the chemical potential $\mu$.

Having solved the frequency renormalization $\Sigma_\omega$, one can
evaluate the pair susceptibility in the gap equation
Eq.~(\ref{eq:gap-imp}),
\begin{eqnarray}
\lefteqn{\chi(0) = \sum_K \frac{\phik^2}{(i\tilde{\omega}-\ek) (
  i\tilde{\underline{\omega}}-\ek) + \Delta^2_\mb{k}} } \nonumber\\
&& = \mbox{Im} \sumk \int_0^\infty \frac{d\omega}{\pi}
  \frac{[1-2f(\omega)]\: \phik^2}
  {(\omega_+ + i\Sigma_+''-\ek) (\omega_- - i\Sigma_-'' - \ek) +
  \Delta^2_\mb{k} }  \,,\nonumber\\
\label{eq:chi-int}
\end{eqnarray}
where $f(x)$ is the Fermi distribution function. It is easy to check
that the gap equation Eq.~(\ref{eq:gap-imp}) does reduce to its clean
counterpart Eq.~(\ref{eq:gap-clean}) as $n_i \rightarrow 0$.

The particle number equation becomes
\begin{equation}
n = 2\sumk \int_{-\infty}^\infty \frac{d\omega}{2\pi}
A(\mb{k},\omega)f(\omega) = -2\, \mathrm{Im} \int_{-\infty}^\infty
\frac{d\omega}{\pi} \dbar{G}^R_\omega f(\omega)\,.
\label{eq:number}
\end{equation}

The real and imaginary parts of $\chi(Q)$ are given respectively by
\begin{subequations}
\begin{eqnarray}
\lefteqn{\chi^\prime(\Omega+i 0^+, \mb{q})}\nonumber\\
&&{} = \mbox{Im} \sumk \int_{-\infty}^\infty \frac{d\omega}{2\pi} \Big\{
  G^R(\omega,\mb{k}) G^R_0(\Omega-\omega, \mb{q-k})\nonumber\\&&{} 
\times \left[ f(\omega-\Omega)-f(\omega) \right]  \nonumber\\
&&{} +  G^R(\omega,\mb{k})  G^A_0(\Omega-\omega, \mb{q-k}) 
\nonumber\\&&{} \times
  \left[1-f(\omega) - f(\omega-\Omega)\right]\Big\} \phikq^2 \,,
\end{eqnarray}
and
\begin{eqnarray}
\lefteqn{\chi''(\Omega+i 0^+, \mb{q})}\nonumber\\
&&{} = - \sumk \int_{-\infty}^\infty \frac{d\omega}{2\pi} 
\mathrm{Im}\, G^R(\omega,\mb{k})
A_0(\Omega-\omega, \mb{q-k})\nonumber\\&&{} \times \left[
    f(\Omega-\omega)-f(\omega) \right] \phikq^2 \,,
\end{eqnarray}
\end{subequations}
where $A_0(\omega, \mb{k}) = -2\mbox{Im}\, G^R_0(\omega, \mb{k})$ is the
``bare'' spectral function.

The pseudogap is evaluated via Eq.~(\ref{eq:PG-imp}). To this end, we
expand the inverse $T$-matrix to the order of $\Omega$ and $q^2$ via a
(lengthy but straightforward) Taylor expansion
\begin{eqnarray}
t^{-1}_{pg}(\Omega+ i 0^+, \mb{q}) &=& \chi(\Omega+ i 0^+, \mb{q})
-\chi(0,\mb{0}) \nonumber\\
 &=& (a_0^\prime + ia_0'') \Omega + b^\prime q^2 + ia_1''\Omega^2 \, .
\label{eq:Tpg-exp}
\end{eqnarray}
Here the imaginary part $b''$ vanishes. The term $b^\prime q^2$ should
be understood as $b^\prime_\parallel q_\parallel^2 + b^\prime_\perp
q_\perp^2$ for a quasi-two dimensional square lattice. We keep the
imaginary part up to the $\Omega^2$ order.  Substituting
Eq.~(\ref{eq:Tpg-exp}) into Eq.~(\ref{eq:PG-imp}), we obtain
\begin{equation}
  \Delta_{pg}^2 = \sum_\mb{q} \int_{-\infty}^\infty \frac{d\Omega}{\pi} 
\frac{(a_0'' +
    a_1''\Omega)\Omega} {(a_0^\prime \Omega + b^\prime q^2)^2 + (a_0'' +
    a_1''\Omega)^2\Omega^2} b(\Omega) \,,
\label{eq:PG}
\end{equation}
where $b(x)$ is the Bose distribution function.

Numerical solutions confirm that the coefficients $a^\prime_0$ and
$b^\prime$ have very weak $T$ dependence at low $T$. Therefore, in a
three-dimensional (3D) system, we have $\Delta_{pg}^2 \sim T^{3/2}$ at
low $T$. As the system dimensionality approaches two, the exponent
decreases from $3/2$ to $1$. However, in most physical systems, e.g.,
the cuprates, this exponent is close to $3/2$. The product $n_p\equiv
a_0^\prime \Delta_{pg}^2$ roughly measures the density of incoherent
pairs.

The gap equation (\ref{eq:gap-imp}) [together with
Eq.~(\ref{eq:chi-int})], the fermion number equation (\ref{eq:number}),
and the pseudogap equation (\ref{eq:PG}) form a closed set, which will
be solved self-consistently for $T_c$, $\mu$, and gaps at and below
$T_c$.  For given parameters $T_c$, $\mu$, and gaps, we can calculate
the frequency renormalization, $\Sigma_\omega$, and then solve the three
equations.  A equation solver is then used to search for the solution
for these parameters. The momentum sum is carried out using integrals.
Very densely populated data points of $\omega$ are used automatically
where $\Sigma_\omega$ and/or $\dbar{G}_\omega$ change sharply. A smooth,
parabolic interpolation scheme is used in the integration with respect
to $\omega$.  The relative error of the solutions is less than
$1.0\times 10^{-5}$, and the equations are satisfied with a relative
error on both sides less than $1.0\times 10^{-7}$. In this way, our
numerical results are much more precise than those calculated on a
finite size lattice.

The solutions of these equations can be used to calculate the superfluid
density $n_s/m$.  Without giving much details, we state here that the
impurity dressing of the current vertex for the (short-coherence-length)
cuprate superconductors does not lead to considerable contributions for
$s$-wave isotropic scattering and the long wavelength $q\rightarrow 0$
limit. The expression for the in-plane $n_s/m$ is given formally by the
formula for a clean system (before the Matsubara summation is carried
out), as in Ref.~\onlinecite{Kosztin-PRB00},
\begin{equation}
\frac{n_s}{m} = \frac{n}{m} + P(0) \,, 
\end{equation}
where the in-plane current-current correlation function
$P_{ij}(Q)=P(Q)\delta_{ij}$ can be simply derived from Eqs.~(31) and
(32) of Ref.~\onlinecite{Kosztin-PRB00}. The result is given by
\begin{eqnarray}
P(Q) &=& \sum_K G(K)G(K-Q)\bigg\{ \Big[1 + (\Delta_{sc}^2 -
\Delta_{pg}^2)\varphi_\mb{k} \varphi_\mb{k-q}\nonumber\\
&&{}\times 
 G_0(-K) G_0 (Q-K) \Big]
\left(\frac{\partial \epsilon_{\mb{k}-\mb{q}/2}}{\partial
    \mb{k}_\parallel}\right)^2 \nonumber\\
&&{} - \Delta_{pg}^2   G_0(K) 
\frac{\partial \epsilon_{\mb{k}-\mb{q}/2}}{\partial \mb{k}_\parallel} \cdot  
\frac{\partial \varphi^2_\mb{k}}{\partial \mb{k}_\parallel} \bigg\}  \,.
\end{eqnarray}
Using spectral representation and after lengthy but straightforward
derivation, we obtain
\begin{eqnarray}
\frac{n_s}{m} &=& 4 \Delta_{sc}^2 \sum_\mb{k} \int_{-\infty}^\infty
\frac{d\omega}{2\pi}\; \mathrm{Im} \left[ \left(\tilde{F}^A(\omega,
    \mb{k})\right)^2
  (\vec{\nabla}\epsilon_\mb{k})^2 \varphi_\mb{k}^2 \right.
\nonumber\\
&&{} + \left. \frac{1}{2} G^A(\omega, \mb{k})\tilde{F}^A(\omega, \mb{k})
  \vec{\nabla}\epsilon_\mb{k} \cdot \vec{\nabla}\varphi^2_\mb{k}\right]
f(\omega) \,, 
\label{eq:Ns}
\end{eqnarray}
where 
\begin{equation}
\tilde{F}(K) = G(K) G_0(-K) =\frac{1}{(i\tilde{\omega}-\ek)
(i\tilde{\underline{\omega}} -\ek) + \Delta^2_\mb{k}}\,,
\end{equation}
which differs from $F(K)$ by a factor $\Delta_\mb{k}$.

As in the clean system, Eq.~(\ref{eq:Ns}) differs from its BCS
counterpart $(n_s/m)_{BCS}$ only by the overall prefactor,
$\Delta_{sc}^2$,
\begin{equation}
\frac{n_s}{m\;} = \frac{\Delta_{sc}^2}{\Delta^2}
\left(\frac{n_s}{m}\right)_{BCS}\;.
\end{equation}
For $d$-wave superconductors, $(n_s/m)_{BCS} \sim A-BT$ or $A-BT^2$ at
very low $T$, depending on whether the system is clean or dirty. Bearing
in mind that $\Delta_{sc}^2/ \Delta^2 = 1 -\Delta_{pg}^2/\Delta^2 \sim
1-C^\prime T^{3/2}$, we have
\begin{eqnarray}
\frac{n_s}{m} &\sim & A-BT-CT^{3/2}\;, \qquad \mbox{(clean)},\nonumber\\
&\sim& A-BT^2 -CT^{3/2},  \qquad \mbox{(dirty)},
\label{eq:Ns-power}
\end{eqnarray}
at very low $T$. In other words, pair excitations lead to a new
$T^{3/2}$ term in the low $T$ superfluid density.

\subsection{Renormalization of the frequency by impurity scattering and
  the density of states}

\begin{figure}
\centerline{\includegraphics[width=3.3in, clip]{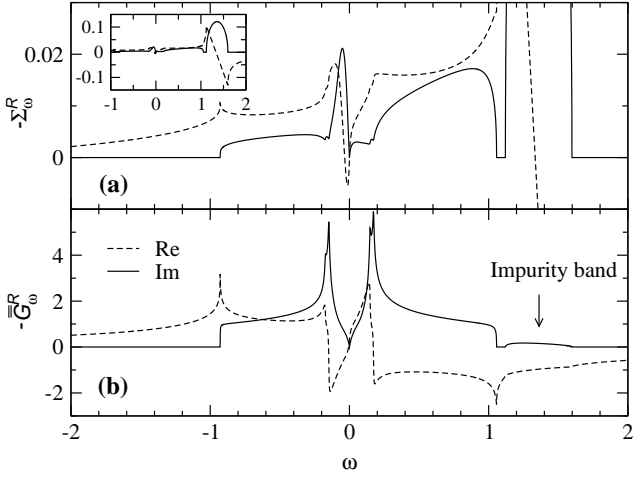}}
\caption{Example of (a) the frequency renormalization $-\Sigma_\omega^R$
  and (b) impurity average of Green's function $-\dbar{G}^R_\omega$ for
  a $d$-wave superconductor. The dashed and solid lines denote the real
  and imaginary parts, respectively. The full impurity band for
  $-\Sigma^R_\omega$ is shown in the inset. The parameters used are
  $\mu=0.92$, $t_\perp/t_\parallel=0.01$, $n_i = 0.02$, $u=1$,
  $\Delta=0.08$. The energy unit is $4t_\parallel$, the half bandwidth.
  }
\label{fig:DOS}
\end{figure}

Except in the Born limit, for $u \gtrsim 4t_\parallel$, impurity
scattering usually introduces a sharp resonance close to the Fermi level
in the frequency renormalization for a $d$-wave superconductor.  In
addition, it induces an impurity band outside the main particle band.
Both the low energy resonance and the high energy impurity band arise
from the vanishing of the real part of the denominator of the impurity
$T$-matrix, Eq.~(\ref{eq:Tfinal}), while the imaginary part is small. In
Fig.~\ref{fig:DOS} we show an example of (a) the retarded, impurity
induced renormalization of the frequency,
$-\Sigma^R_\omega=\tilde{\omega}^R-\omega$ and (b) the corresponding
impurity average of the single-particle Green's function,
$-\dbar{G}^R_\omega$, which is related to the density of states by
$N(\omega)=-2\, \mbox{Im}\,\dbar{G}^R_\omega $.  The curves for
$-\Sigma^R_\omega$ corresponding to the impurity band are replotted in
the inset, to show the strong renormalization of the frequency inside
the impurity band.  (The curves for $-\Sigma^\prime_\omega$ has been
offset so that $-\Sigma^\prime_{\omega=0}=0$. This offset is compensated
by a shift in the chemical potential $\mu$.) The van Hove singularity
and its mirror image via particle-hole mixing are clearly seen in the
density of states, and are also reflected in $-\Sigma^R_\omega$ as the
small kinks in Fig.~\ref{fig:DOS}(a). The real part
$-\Sigma^\prime_\omega$ is usually neglected in most non-self-consistent
calculations.\cite{Hirschfeld88} It is evident, however, that it has a
very rich structure, and, in general, cannot be set to zero in any
self-consistent calculations. This conclusion holds even in the presence
of exact particle-hole symmetry, as can be easily told from
Eq.~(\ref{eq:Tfinal}).

For $u<0$, the low energy resonance in $-\mbox{Im}\,\Sigma^R_\omega$
will appear on the positive energy side in Fig.~\ref{fig:DOS}(a).
Regardless of the sign of $u$, the resonance peak will become sharper as
$n_i$ decreases and as $|u|$ increases. For larger $|u|$, the resonant
frequency will be closer to $\omega=0$, where
$-\mbox{Im}\,\dbar{G}^R_\omega$ is small because of the $d$-wave
symmetry; A smaller $n_i$ further reduces
$-\mbox{Im}\,\dbar{G}^R_\omega$. Both factors help minimize the
imaginary part of the denominator of Eq.~(\ref{eq:Tfinal}) and, thus,
lead to a stronger resonance.  It should be emphasized that a resonance
peak does not show up in $-\mbox{Im}\,\dbar{G}^R_\omega$ since the
resonance in $-\mbox{Im}\,\Sigma^R_\omega$ requires that
$-\mbox{Im}\,\dbar{G}^R_\omega$ be small at the resonance point. 

\begin{figure}
\centerline{\includegraphics[width=3.3in, clip]{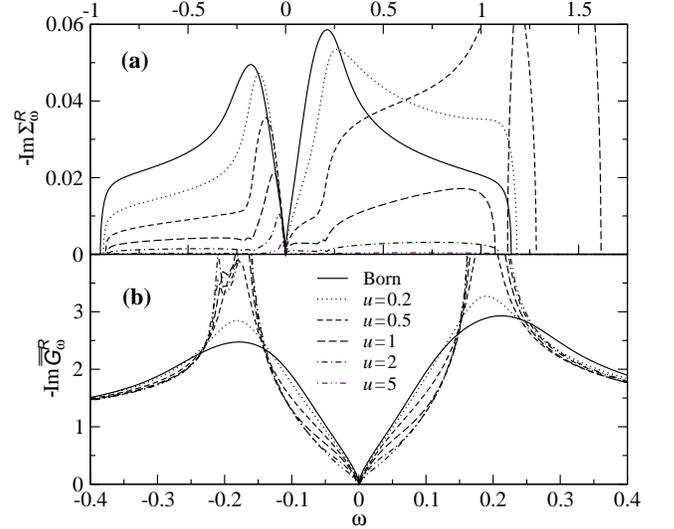}}
\caption{Evolution of (a) the frequency renormalization $-\mbox{Im}\, 
  \Sigma_\omega^R$ and (b) impurity average of Green's function
  $-\mbox{Im}\,\dbar{G}^R_\omega$ with $u$ for a $d$-wave superconductor
  at fixed $\gamma=n_i u^2 = 0.02$. A resonance develops in
  $-\mbox{Im}\, \Sigma_\omega^R$ as $u$ deviates from the Born limit.
  The weak scattering (Born limit) is more effective in filling in the
  DOS within the gap. The parameters used are $\mu=0.9$,
  $t_\perp/t_\parallel=0.01$, $\Delta=0.0945$. }
\label{fig:gamma-u}
\end{figure}

The location of the impurity band is sensitive to the sign and strength
of impurity scattering. For $u<0$, the impurity band on the negative
energy (left) side of the plot.  As $|u|$ gets smaller, the impurity
band merges with the main band; as $|u|$ gets larger, it moves farther
away, with a much stronger renormalization of $\omega$.  For large
$|u|$, the spectral weight under the impurity band in
Fig.~\ref{fig:DOS}(b) is given by $2n_i$, and the weight in the main
band is reduced to $2(1-n_i)$.  This leads to a dramatic chemical
potential shift as a function of the impurity concentration $n_i$ (as
well as $u$). For $u<0$, the impurity band will always be filled, so
that increasing $n_i$ pushes the system farther away from the
particle-hole symmetry. For $u>0$, on the contrary, the impurity band is
empty, and the system becomes more particle-hole symmetrical as $n_i$
increases from 0, and reaches the particle-hole symmetry (in the main
band) at $n_i=1-n$. This fact implies that P\'epin and Lee's assumption
of an exact particle-hole symmetry is not justified so that their
prediction of a diverging $N(\omega)$ as $\omega \rightarrow 0$ is
unlikely to be observed experimentally. It should be mentioned that the
appearance of the impurity band has not been shown in the literature,
largely because most authors concentrate on the low energy part of the
spectra only, and do not solve for the full spectrum of the
renormalization of $\omega$ self-consistently.

\begin{figure}
\centerline{\includegraphics[width=3.3in, clip]{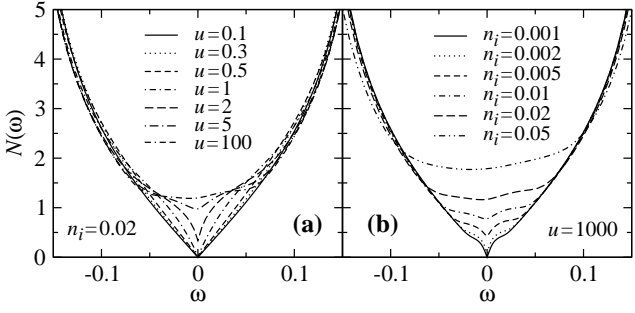}}
\caption{Evolution of the DOS 
  $N(\omega)$ for a $d$-wave superconductor (a)
  with $u$ at $n_i=0.02$ and (b) with $n_i$ in the unitary limit $u =
  1000$.  There is a dip at $\omega=0$ for small $n_i$ or small $u$.
  The parameters used are $\mu=0.9$, $t_\perp/t_\parallel=0.01$,
  $\Delta=0.0945$. }
\label{fig:-u-ni}
\end{figure}
 
In the Born limit, only the product $\gamma=n_i u^2$ is a meaningful
parameter, not $n_i$ or $u$ separately. In Fig.~\ref{fig:gamma-u}, we
plot (a) the frequency renormalization $-\mbox{Im}\, \Sigma_\omega^R$
and (b) the impurity average of Green's function
$-\mbox{Im}\,\dbar{G}^R_\omega$ as a function of $\omega$ for various
values of $u$ but with a fixed $\gamma=0.02$.  (Note: when $u$ is small,
this requires an unphysically large $n_i$.)  In the Born limit, these
two quantities are identical up to a constant coefficient. As $u$
increases, a resonance develops in $-\mbox{Im}\, \Sigma_\omega^R$ at
small $\omega$. $-\mbox{Im}\, \Sigma_\omega^R$ and
$-\mbox{Im}\,\dbar{G}^R_\omega$ become very different. And a impurity
band develops gradually ($u=0.2$ and $u=0.5$), until it splits from the
main band ($u=1$). At fixed $\gamma$, the Born limit is more effective
in filling in the DOS in the mid-range of $\omega$ within the gap and
smearing out the coherence quasiparticle peaks, whereas the large $u$
limit is more effective in filling in the DOS in the vicinity of
$\omega=0$ but keeping the quasiparticle peaks largely unchanged. In
addition, the main band becomes narrower at large $u$ than that in the
clean system or the Born limit, so that part of the spectral weight has
now been transferred to the impurity band. Also note that the DOS at
$\omega=0$ is essentially zero in Fig.~\ref{fig:gamma-u}(b) because
$n_i$ is very small when $u$ becomes large for the current choice of
$\gamma=0.02$.

The effects of the scattering strength $u$ and the impurity density
$n_i$ on the DOS are shown in Fig.~\ref{fig:-u-ni}(a) and (b),
respectively. For the effect of $u$ in Fig.~\ref{fig:-u-ni}(a), we
choose an intermediate $n_i =0.02$. And for the effect of $n_i$, we
focus on the unitary limit, and choose $u=1000$. There is a dip in the
DOS at $\omega=0$ in both the small $u$ and small $n_i$ cases, mimicking
a fractional power law dependence on $\omega$. At higher $n_i$ and
higher $u$, the DOS is filled in mainly at small $\omega$.

\begin{figure}
\centerline{\includegraphics[width=3.37in, clip]{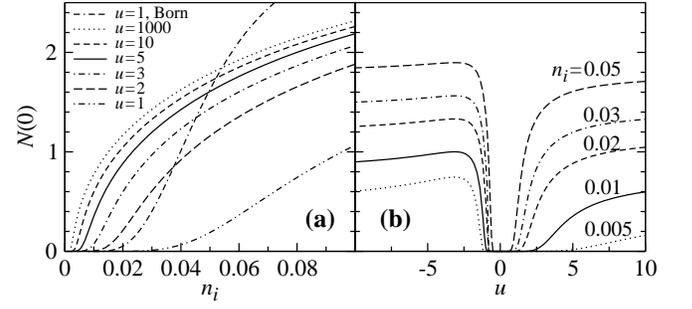}}
\caption{Residue DOS $N(0)$ at the Fermi level as a function of (a)
  $n_i$ for $u=1000$, 10, 5, 3, 2, and 1, and of (b) $u$ for
  $n_i=0.005$, 0.01, 0.02, 0.03, and 0.05.
  Also plotted in (a) is $N(0)$ as a function of $\gamma (=n_i)$ in the
  Born limit.
  The parameters used are $\mu=0.9$, $t_\perp/t_\parallel=0.01$,
  $\Delta=0.0945$. }
\label{fig:-u_ni}
\end{figure}

Shown in Fig.~\ref{fig:-u_ni} are the residue DOS at the Fermi level,
$N(0)$, as a function of (a) the impurity concentration $n_i$ for
different values of $u$ from the unitary limit $u=1000$ through $u=10$,
5, 3, 2, down to 1, and (b) as a function of the scattering strength $u$
for $n_i=0.005$, 0.01, 0.02, 0.03, and 0.05. Figure \ref{fig:-u_ni}(a)
indicates that below certain ``critical'' value of $n_i$, $N(0)$ remains
essentially zero. This behavior is also implied by the presence of the
dip at small $n_i$ in Fig.~\ref{fig:-u-ni}(b).  The ``critical'' value
for $n_i$ in Fig.~\ref{fig:-u_ni}(a) is clearly scattering strength
dependent. The smaller $u$, the larger this value. A replot (not shown)
of these curves in terms of $\log_{10}N(0)$ as a function of $1/n_i$
reveals that for small $n_i$, $N(0)$ vanishes exponentially as
$e^{-A/n_i}$, where $A$ is a constant. For comparison, we also show in
Fig.~\ref{fig:-u_ni}(a) the Born limit as a function of $\gamma (=n_i)$.
As one may expect, the Born limit is rather different from the rest, since
it is equivalent to a very small $u<<1$ and unphysically large $n_i$. A
similar activation behavior of $N(0)$ as a function of $u$ is seen in
Fig.~\ref{fig:-u_ni}(b), where the ``critical'' value for $u$ is
strongly $n_i$ dependent. The asymmetry between positive and negative
$u$ reflects the particle-hole asymmetry at $\mu=0.9$. It should be
noted that it is not realistic to vary $u$ continuously in experiment.


An earlier experiment by Ishida \textit{et al.}\cite{Ishida} suggests
that $N(0)$ varies as $n_i^{1/2}$.  In our calculations, however, $N(0)$
does not follow a simple power law as a function of $n_i$.  The curve
for $u=1000$ in Fig.~\ref{fig:-u_ni}(a) fits perfectly with $a
n_i^\alpha -b$, with $\alpha\approx 0.175$, for $0.002 < n_i < 0.05$.
The damping of the zero frequency (not shown), $-\mbox{Im}
\Sigma^R_{\omega=0}$, also fits this functional form very well, with
$\alpha\approx 0.61$. The exponents are different for different values
of $u$.  Our calculation for the $n_i$ dependence of $N(0)$ is
consistent with the result of Fehrenbacher\cite{Fehrenbacher} in that
$N(0, n_i)$ it is strongly $u$ dependent. However, it does not seem likely
that the simple power law $N(0) \sim n_i^{1/2}$ may be obtained in an
accurate experimental measurement. Further experiments are needed to
double check this relationship between $N(0)$ and $n_i$.

From Figs.~\ref{fig:DOS}-\ref{fig:-u_ni}, we conclude that for very
small $n_i$ and $u$, the zero frequency DOS $N(0)$ is exponentially
small.  At high $n_i$, $N(0)$ is finite in both the Born and the unitary
limit.  However, for certain intermediate values of $n_i$ and $u$ [e.g.,
$u\sim 1$ in Fig.~\ref{fig:-u-ni}(a), and $n_i \lesssim 0.002$ in
Fig.~\ref{fig:-u-ni}(b)], $N(\omega)$ vanishes with (very small)
$\omega$ according to some fractional power $\omega^\alpha$ where
$\alpha < 1$.  We see neither the universal power laws for $N(\omega)$
predicted by Senthil and Fisher,\cite{Fisher} nor the divergent DOS
predicted by P\'epin and Lee\cite{Lee} and others.\cite{Bocquet}

When the values of $u$ and $n_i$ are such that $N(\omega) \sim
\omega^\alpha$ with $\alpha < 1$, one may expect to see a fractional low
$T$ power law in the superfluid density. However, such a power law is
not robust as it is sensitive to the impurity density for a given type
of impurity. The situation with a negative $u$ is similar to
Fig.~\ref{fig:-u-ni}.

For given chemical potential $\mu$ and the total excitation gap
$\Delta$, the calculation of the frequency renormalization with
impurities does not necessarily involve the concept of the pseudogap. It
is essentially the ``self-consistent'' impurity $T$-matrix treatment by
Hirschfeld \textit{et al.}\cite{Hirschfeld88} except that we now have to
solve for both the real and imaginary parts of $\Sigma_\omega^R$
simultaneously in a self-consistent fashion.  Our numerical results
agree with existing calculations in the literature.

Finally, we emphasize the difference between the self-consistent
impurity $T$-matrix treatment of the one-impurity problem
\cite{Balatsky,Hirschfeld00} and the current many-impurity averaging.
For the former case, the impurity $T$-matrix will be given by
Eq.~(\ref{eq:Tw}) but with $\dbar{G}_\omega$ replaced by
$\dbar{G}^0_\omega$, i.e., the impurity average of the \textit{clean}
$G^0(K)$.  As a consequence, the position of the poles of $T_\omega$ is
independent of the renormalized DOS, and, therefore, a resonance peak
may exist at low $\omega$ in the DOS,\cite{Balatsky} whereas it cannot
in the current many-impurity case.

%

\begin{figure}
\centerline{\includegraphics[width=3.3in, clip]{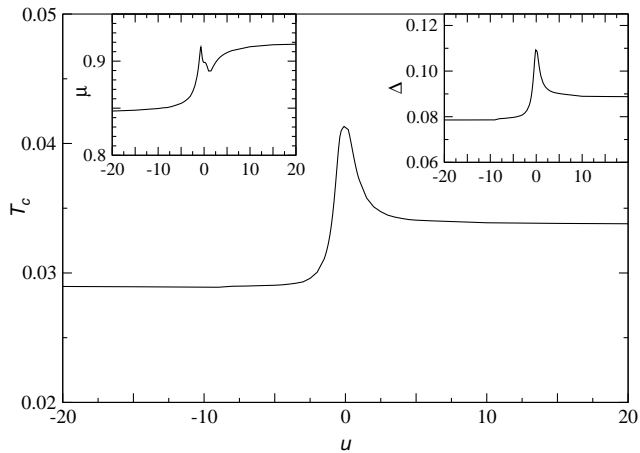}}
\caption{Behavior of $T_c$, $\mu$ (left inset), and $\Delta$ (right inset) 
  as a function of the impurity scattering strength $u$ at $n=0.85$,
  $t_\perp/t_\parallel=0.01$, $-g/4t_\parallel=0.5$ and $n_i = 0.05$.  }
\label{fig:Tc-u}
\end{figure}

\subsection{Effects of the impurity scattering on $T_c$ and the pseudogap}
\label{subsec:Tc}

In this section, we study the influence of impurity scattering on the
behavior of $T_c$ and the pseudogap $\Delta_{pg}$ as a function of the
coupling strength as well as the hole doping concentration.

First, we study the effect of the scattering strength $u$ and whether it
is repulsive ($u>0$) or attractive ($u<0$). In Fig.~\ref{fig:Tc-u}, we
plot $T_c$ as a function of $u$, for a pseudogapped $d$-wave
superconductor with $n_i=0.05$. The corresponding curves for $\mu$ and
$\Delta=\Delta_{pg}(T_c)$ are shown in the upper left and upper right
insets, respectively.  All three quantities, $T_c$, $\mu$, and $\Delta$,
vary with $u$. For either sign of $u$, both $T_c$ and $\Delta$ are
suppressed by increasing $|u|$. It should be emphasized that the
chemical potential $\mu$ in the two (large $\pm u$) unitary limits does
not meet, nor does $T_c$ or $\Delta$. This is because a (large)
negative $u$ creates a \textit{filled} impurity band below the main
band, and is effective in bringing down the chemical potential, whereas
a positive $u$ creates an \textit{empty} impurity band above the main
band, and tends to raise $\mu$ toward the particle-hole symmetrical
point, $\mu=1$. This result cannot and has not been observed in previous,
non-self-consistent calculations where the real part of frequency
renormalization is set to zero. 

\begin{figure}
\centerline{\includegraphics[width=3.3in, clip]{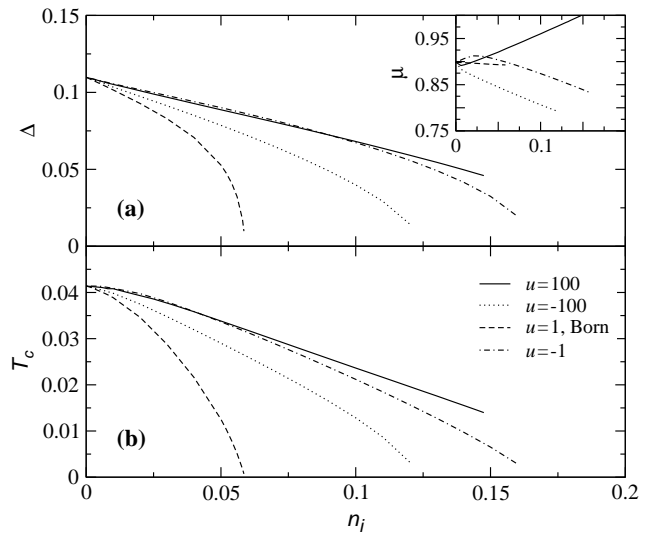}}
\caption{Evolution of (a) $\Delta_{pg}(T_c)$, (b) $T_c$, and $\mu$ (inset), 
  as a function of the impurity concentration $n_i$ for both positive
  ($u=100$) and negative ($u=-100$) scattering strength in the unitary
  limit, the Born limit ($u=1$, $\gamma=n_i$), and intermediate $u=-1$.
  Here $n = 0.85$, $t_\perp/t_\parallel=0.01$, and
  $-g/4t_\parallel=0.5$.  }
\label{fig:Tc-ni}
\end{figure}

In Fig.~\ref{fig:Tc-ni}, we compare the effect of the impurity
concentration for different scattering strengths: the Born limit, both
unitary limits ($u=\pm 100$), and intermediate $u=-1$. Both (b) $T_c$
and (a) $\Delta_{pg}(T_c)$ are suppressed by increasing impurity
density. This is natural in a model where the pseudogap originates from
incoherent pair excitations. As will be seen below, $\Delta_{pg}(T_c)$
is suppressed mainly because $T_c$ is lowered. Except in the Born limit,
the chemical potential $\mu$ is fairly sensitive to $n_i$, as shown in
the inset. It is clear that the scattering in the Born limit is the most
effective in suppressing $T_c$.  In comparison with
experiment,\cite{Franz} calculations at the AG level (i.e., the Born
limit) tend to overestimate the $T_c$ suppression by as much as a factor
of 2. This is in good agreement with the current result in the unitary
limit.
At large $n_i$ for large positive $u$, the system is driven to the
particle-hole symmetrical point, where the effective pair mass changes
sign. It is usually hard to suppress $T_c$ by pairing at the
particle-hole symmetrical point, as indicated by the solid curve in the
lower panel. In fact, exactly at this point, the linear $\Omega$ term
$a_0^\prime$ in the inverse $T$ matrix expansion vanishes, so that one
needs to go beyond the current approximation and expand up to the
$\Omega^2$ term.  Large negative $u=-100$ is more effective in
suppressing $T_c$ and $\Delta$ than intermediate negative $u=-1$, in
agreement with Fig.~\ref{fig:Tc-u} and the DOS shown in
Fig.~\ref{fig:-u-ni}(a) and Fig.~\ref{fig:-u_ni}.  

\begin{figure}
\centerline{\includegraphics[width=3.3in, clip]{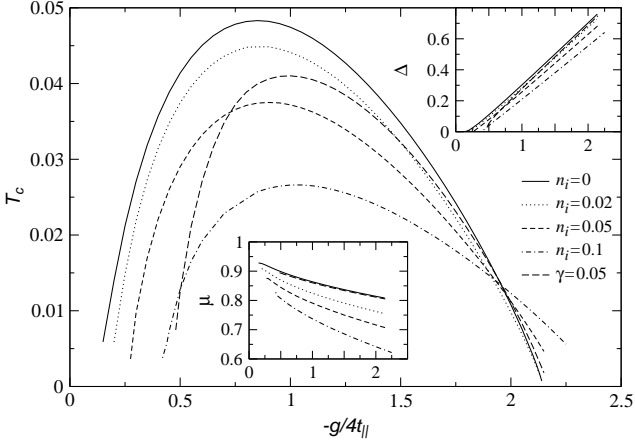}}
\caption{Behavior of $T_c$, $\mu$ (lower inset), and $\Delta$ (upper
  inset) as a function of $-g/4t_\parallel$ in the unitary limit for $u=
  -100$ and $n_i= 0$ (solid curve), 0.02 (dotted), 0.05 (dashed), and
  0.1 (dot-dashed). For comparison, curves for $\gamma =0.05$ in the
  Born limit are also plotted (long-dashed). Impurity scattering in the
  Born limit is more effective in suppressing $T_c$ at relatively weak
  coupling. The parameters are $n=0.85$, $t_\perp/t_\parallel=0.01$.  }
\label{fig:Tc-g}
\end{figure}

There is enough evidence that zinc impurities are attractive scatterer
for electrons in the cuprates.\cite{Fehrenbacher} Therefore, we
concentrate ourselves on negative $u$ scattering in the unitary limit.
Plotted in Fig.~\ref{fig:Tc-g} are $T_c$ (main figure), $\mu$ (lower
inset), and $\Delta$ as a function of $g$ for increasing $n_i=0, 0.02,
0.05, 0.1$ with $u=-100$.  Also plotted for comparison are the results
assuming the Born limit with $\gamma=0.05$. Clearly, the Born limit is
more effective in suppressing $T_c$ at relatively weak coupling,
$-g/4t_\parallel \lesssim 0.75$, consistent with Fig.~\ref{fig:Tc-ni}.
Both $T_c$ and $\Delta$ are suppressed continuously with $n_i$.
However, it should be noted that a larger $n_i$ helps $T_c$ to survive a
larger $-g/4t_\parallel$. This is mainly because the filled impurity
band at large $n_i$ pushes the system far away from particle-hole
symmetry (see $\mu$ in the lower inset), and reduces the effective
fermion density, so that the pair mobility is enhanced and the pair mass
does not diverge until a larger $-g/4t_\parallel$ is reached.

\begin{figure}
\centerline{\includegraphics[width=3.3in, clip]{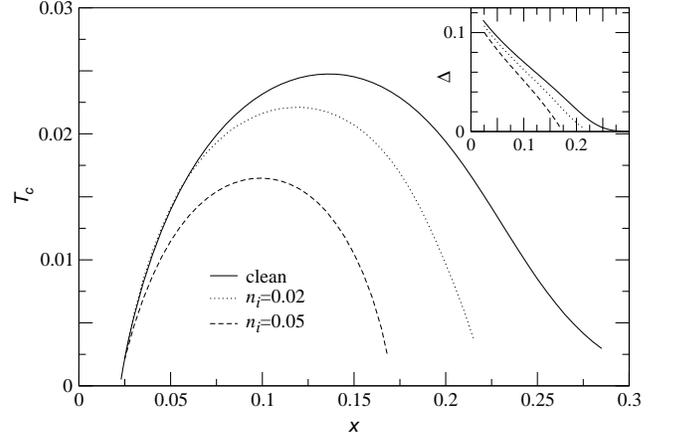}}
\caption{Behavior of $T_c$  and $\Delta$ (inset) 
  as a function of the hole doping concentration $x$ in the unitary
  limit for $u= -100$ and $n_i= 0$ (solid curve), 0.02 (dotted), and
  0.05 (dashed). The parameters are $-g/4t_0=0.047$,
  $t_\perp/t_\parallel=0.003$.  }
\label{fig:Tc-x}
\end{figure}

To make contact with the cuprates, we use the non-double occupancy
condition associated with the Mott insulator physics, as in
Ref.~\onlinecite{Chen-PRL98}, so that the effective hopping integral is
reduced to $t_\parallel(x)\approx t_0 x$, where $x=1-n$ is the hole
concentration, and $t_0 = 0.6$~eV is the hopping integral in the absence
of the on-site Coulomb repulsion. We assume $-g/4t_0 = 0.047$, which is
$x$ independent. Then we can compute $T_c$, $\mu$, and
$\Delta=\Delta_{pg}(T_c)$ as a function of $x$. The result for $T_c$
(main figure) and $\Delta$ (inset) is shown in Fig.~\ref{fig:Tc-x} for
the clean system and $n_i = 0.02$ and $0.05$.  In the overdoped regime,
$T_c$, as well as the small $\Delta$, are strongly suppressed by
impurities.  This provides a natural explanation for the experimental
observation that $T_c$ vanishes abruptly at large $x$; it is well-known
that high crystallinity, clean samples are not available in the extreme
overdoped regime. On the other hand, neither $T_c$ nor $\Delta$ is
strongly suppressed in the highly underdoped regime, where the gap is
too large.  At this point, experimental data in this extreme underdoped
regime are still not available. Our result about the suppression of
$T_c$ and $\Delta$ in the less strongly underdoped regime ($x> 0.1$) are
in agreement with experimental observations\cite{Tallon95,Tallon98} and
other calculations.\cite{Franz}

It should be noted, however, that in our simple model, we do not
consider the fact that disorder or impurities may reduce the
dimensionality of the electron motion and thus suppress $T_c$.
Furthermore, since induced local spin and Kondo effects have been
observed near zinc sites in both zinc-doped
YBCO\cite{MacFarlane,Bobroff99} and zinc-doped
Bi$_2$Sr$_{2-x}$La$_x$CuO$_{6+\delta}$,\cite{Hanaki} this raises an
important question whether zinc can be treated as a nonmagnetic
impurity.

\subsection{Gaps and superfluid density below $T_c$ 
  in the presence of nonmagnetic impurities}

In this subsection, we study the effect of nonmagnetic impurities on the
behavior of the excitation gap $\Delta$, the order parameter
$\Delta_{sc}$, and the pseudogap $\Delta_{pg}$ as well as the chemical
potential $\mu$ as a function of temperature in the superconducting
phase. The numerical solutions for these quantities are then used to
study the temperature dependence of the superfluid density $n_s/m$ at
various impurity levels. We concentrate on the unitary limit, which is
regarded as relevant to the cuprates. To be specific, we use $n=0.85$,
$t_\perp/t_\parallel =0.01$, and $u=-100$ in the calculations
presented below.

\begin{figure}
\centerline{
\includegraphics[width=3.3in, clip]{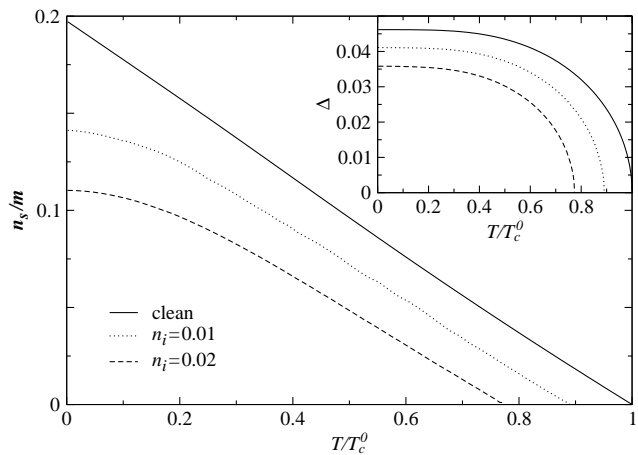}}
\caption{Behavior of the superfluid density $n_s/m$  and the excitation
  gap $\Delta$ (inset) in a $d$-wave BCS superconductor as a function of
  temperature $T/T_c^0$ for impurity concentration $n_i=0$ (clean, solid
  curve), 0.02 (dotted), and 0.05 (dashed) in the unitary limit $u=
  -100$. $T_c^0=0.0416$ is the $T_c$ in the clean system. The parameters
  are $n=0.85$, $-g/4t_\parallel=0.3$, and $t_\perp/t_\parallel=0.01$.
  }
\label{fig:Ns-BCS}
\end{figure}

We first study the impurity effect in the BCS case, without the
complication of the pseudogap. Plotted in Fig.~\ref{fig:Ns-BCS} are the
superfluid density $n_s/m$ (main figure) and the corresponding gap
$\Delta$ in a $d$-wave BCS superconductor as a function of the reduced
temperature $T/T_c^0$ for the clean system (solid curves), impurity
density $n_i=0.02$ (dotted), and $n_i=0.05$ (dashed) at $-g/4t_\parallel
=0.3$. Here $T_c^0=0.0416$ is the $T_c$ in the clean case. As expected,
both $T_c$ and $\Delta(T)$, as well as $n_s/m$, are suppressed by
impurity scattering. In agreement with experiment, the low $T$ normal
fluid density is linear in $T$ in the clean case, and becomes quadratic
in the two dirty cases.

\begin{figure}
\centerline{\includegraphics[width=3.3in, clip]{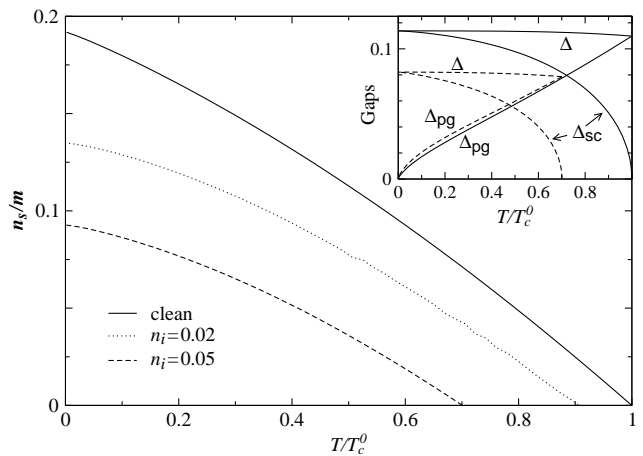}}
\caption{Behavior of the superfluid density $n_s/m$  and the various
  gaps (inset) in a $d$-wave pseudogapped superconductor as a function
  of temperature $T/T_c^0$ for impurity concentration $n_i=0$ (clean,
  solid curve), 0.02 (dotted), and 0.05 (dashed) in the unitary limit
  $u= -100$. Here $T_c^0=0.0414$. The parameters are $n=0.85$,
  $-g/4t_\parallel=0.5$, and $t_\perp/t_\parallel=0.01$.  }
\label{fig:Ns}
\end{figure}

Now we add pseudogap for the underdoped cuprates. We show in
Fig.~\ref{fig:Ns} the temperature dependence of $n_s/m$ (main figure)
and various gaps (inset) in a $d$-wave pseudogapped superconductor for
impurity concentration $n_i=0$ (clean, solid curve), 0.02 (dotted), and
0.05 (dashed) in the unitary limit at $-g/4t_\parallel =0.5$. As the
order parameter develops below $T_c$, the pseudogap decreases with
decreasing $T$. This reflects the fact that the pseudogap in the present
model is a measure of the density of thermally excited pair excitations.
The total gap $\Delta$, the order parameter $\Delta_{sc}$, and the
superfluid density $n_s/m$ are suppressed by increasing $n_i$, similar
to the BCS case above.  However, at given $T<T_c$, the pseudogap
$\Delta_{pg}$ remains roughly unchanged.  Furthermore, the low $T$ power
law for the superfluid density is different from the BCS case, as
predicted in Eq.~(\ref{eq:Ns-power}). It is now given by $T+T^{3/2}$ and
$T^2 + T^{3/2}$ for the clean and dirty cases, respectively. Due to the
presence of the $T^{3/2}$ term, the low $T$ portion of the curves for
$n_i =0.02$ and 0.05 are clearly not as flat as in
Fig.~\ref{fig:Ns-BCS}.  Nevertheless, it may be difficult to distinguish
experimentally $T^2+T^{3/2}$ from a pure $T^2$ power law. This $T^{3/2}$
contribution of the pair excitations has been used to explain
successfully\cite{Chen-PRL98} the quasi-universal behavior of the
normalized superfluid density $n_s(T)/n_s(0)$ as a function of $T/T_c$.
We have also found\cite{Chen-PC,Carrington} preliminary experimental
support for this $T^{3/2}$ term in low $T$ penetration depth measurement
in the cuprates as well as organic superconductors. Systematic
experiments are needed to further verify this power law prediction.

\begin{figure}
\centerline{\includegraphics[width=3.3in, clip]{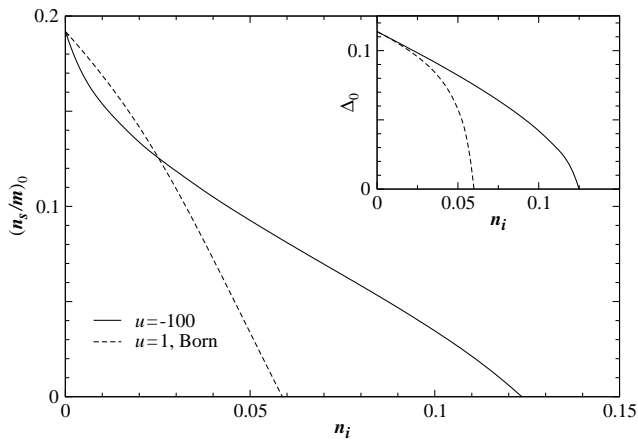}}
\caption{Zero temperature superfluid density $(n_s/m)_0$ and the gap
  $\Delta_0$ (inset) in a $d$-wave superconductor as a function of
  impurity concentration $n_i$ in the unitary limit ($u= -100$, solid
  line) and in the Born limit ($\gamma=n_i$, dashed line).  The
  parameters are $n=0.85$, $-g/4t_\parallel=0.5$, and
  $t_\perp/t_\parallel=0.01$.  }
\label{fig:Ns0}
\end{figure}

A careful look at the values of the zero temperature superfluid density
$(n_s/m)_0$ for different values of $n_i$ in both Fig.~\ref{fig:Ns-BCS}
and Fig.~\ref{fig:Ns} suggests that in the unitary limit, $(n_s/m)_0$
drops faster with $n_i$ when $n_i$ is still small. This is manifested in
a systematic study of $(n_s/m)_0$ as a function of $n_i$, as shown in
Fig.~\ref{fig:Ns0}, with $u=-100$ (solid curve). This behavior has been
observed experimentally.\cite{Bernhard} Also plotted in the inset is the
corresponding zero temperature gap $\Delta_0$ versus $n_i$. Clearly, the
slope $d\frac{n_s}{m}/dn_i$ is much steeper as $n_i$ approaches zero,
very different from the behavior of $\Delta_0$. This demonstrates that
$n_s/m$ is influenced more by the DOS than by the gap size. A very small
amount of impurities may strongly suppress $(n_s/m)_0$. This conclusion
is significant in data analysis of the penetration depth measurement,
especially when the $T=0$ value of penetration depth is not measured
directly.\cite{Hardy} For comparison, we also plot the corresponding
curves in the Born limit. While the gap is suppressed faster, in
contrast to unitary case, the slope $d\frac{n_s}{m}/dn_i$ is smaller for
smaller $n_i$.

\section{Discussion}

In Sec.~\ref{sec:theory}, we have used the approximation
Eq.~(\ref{eq:approx}) to bring the single-particle self-energy and thus
the gap equation into a BCS-like form. This approximation derives from
the divergence of the $T$ matrix $t_{pg}(Q)$ as $Q\rightarrow 0$, which
is the pairing instability condition. The spirit of this approximation
is to ``put'' the incoherent, excited pairs into the condensate, by
setting $Q=0$. The contribution of these pseudo Cooper pairs to the
single-particle excitation gap is calculated via Eqs.~(\ref{eq:PG-imp})
and (\ref{eq:PG}), weighted by the Bose function. Therefore, the
incoherent pairs and the zero momentum condensate are not distinguished
from each other in terms of the single particle self-energy, as they add
up to a total excitation gap. However, they are distinct when phase
sensitive quantities are involved, e.g., in the calculation of $T_c$
and of the superfluid density.

With this approximation, there is a close analogy between the Feynman
diagrams in the current pairing fluctuation theory in
Sec.~\ref{sec:theory} and its BCS counterpart in Appendix \ref{App:BCS}.
When the finite momentum pair propagators are removed (or ``pushed into
the condensate'') from Fig.~\ref{diag-G1} through
Fig.~\ref{diag-Tmatrix}, these diagrams will become their BCS
counterpart in Fig.~\ref{diag:BCS-AG} through
Fig.~\ref{diag:BCS-Tmatrix}. (The diagram for BCS pairing vertex is not
shown in Appendix \ref{App:BCS}).

This approximation is in general good when the gap is large in
comparison with $T$, and when the impurity concentration is low. When the
gap is small, the contribution of the incoherent pair excitations is
usually small, and does not have a strong effect on $T_c$,
so that $T_c$ is roughly determined by its BCS mean-field solution. When
$n_i$ is large, the fermionic frequency renormalization is strong, and
the pair dispersion also becomes highly damped. In this case,
approximation Eq.~(\ref{eq:approx}) may not be quantitatively accurate.

Even without the complication of impurities, incoherent pairs are not
expected to deplete completely the spectral weight within the two
quasiparticle peaks of the spectral function.\cite{Chen-PRB01,Maly1} This,
however, cannot be captured by the approximation Eq.~(\ref{eq:approx}).
Unfortunately, we still do not know yet how to solve the Dyson's
equations without this approximation due to technical difficulties.

Another simplification comes from the $d$-wave symmetry of the cuprate
superconductors under study. Although we have kept the theoretical
formalism general for both $s$- and $d$-wave in Sec.~\ref{sec:theory},
the pairing vertex renormalization drops out when we finally carry out
numerical calculations for $d$-wave superconductors. For $s$-wave
superconductors, one would have to include self-consistently one more
complex equation for the renormalization of $\Delta_\mb{k}$, when
solving for the renormalization of $\omega$. And the equations
(\ref{eq:Tw}) and (\ref{eq:TDelta}) also look much more complicated than
Eq.~(\ref{eq:Tfinal}). Nevertheless, since there is no node in the
excitation gap for $s$-wave, the numerics is expected to run faster.

It is well-known that for $d$-wave superconductors, the Anderson's
theorem\cite{Anderson59} breaks down.\cite{Gorkov} For Anderson's theorem
to hold, it requires that the gap and the frequency are renormalized in
exactly the same fashion.  This condition can be satisfied
(approximately) only in weak coupling, isotropic BCS $s$-wave
superconductors, for which the real part of the frequency
renormalization is negligible. Since the frequency $\omega$ is a scalar,
this condition is violated when the gap $\Delta_\mb{k}$ have any
anisotropic dependence on $\textbf{k}$.  Furthermore, when the gap is
considerably large in comparison with the band width so that the upper
limit of the energy integral cannot be extended to infinity, this
condition will not be satisfied, either. In both cases, $T_c$ will be
suppressed.

\section{Conclusions}

In this paper, we extend the pairing fluction theory to superconductors
in the presence of non-magnetic impurities. Both the pairing and
impurity $T$-matrices are included and treated self-consistently. We
obtain a set of three equations for ($T_c$, $\mu$, $\Delta(T_c)$) or
($\mu$, $\Delta_{sc}$, $\Delta_{pg}$) at $T<T_c$, with the complex
equations for the frequency renormalization. In consequence, we are able
to study the impurity effects on  $T_c$, the order parameter, and the
pseudogap.  In particular, we carry out calculations for $d$-wave
superconductors and apply to the cuprate superconductors. Instead of
studying the physical quantities with all possible combinations of the
parameters $n_i$, $u$, $g$, and $n$, we mainly concentrate on the
negative $u$ unitary limit, which is regarded as relevant to the zinc
impurities in the cuprates.\cite{Fehrenbacher}

Calculations show that in addition to the low energy resonance in the
imaginary part of the renormalized frequency, a considerably large $|u|$
leads to a separate impurity band, with a spectral weight $2n_i$. The
real part of the frequency renormalization, in general, cannot be set to
zero in a self-consistent calculation. The chemical potential varies
with the impurity concentration, so that the assumption of exact
particle-hole symmetry is not justified when one studies the impurity
effects. One consequence of this chemical potential shift is that the
repulsive and attractive unitary scattering limits do not meet as has
been widely assumed in the non-self-consistent treatment in the
literature. Unitary scatterers fill in the DOS mostly in the small
$\omega$ region, whereas Born scatterers do in essentially the whole
range within the gap. At small $n_i$ and/or small $u$, there is a dip at
$\omega=0$ in the DOS, so that $N(\omega)$ vanishes as a fractional
power of $\omega$, which may in turn contribute a fractional power law
for the low $T$ temperature dependence of the penetration depth.

Both $T_c$ and the pseudogap $\Delta_{pg}(T_c)$ are suppressed by
impurities. In this respect, Born scatterers are about twice as
effective as unitary scatterer. Treating zinc impurities as unitary
scatterers explains why the actual $T_c$ suppression is only half that
predicted by calculations at the AG level (i.e., in the Born limit). In
the overdoped regime, the gap is small, and therefore the
superconductivity can be easily destroyed by a small amount of
impurities. In contrast, it takes a larger amount of impurities to
destroy the large excitation gap in the underdoped regime.

The reason $\Delta_{pg}(T_c)$ is suppressed is mainly because $T_c$ is
suppressed. In fact, for a given $T< T_c$, the pseudogap remains roughly
unchanged (actually it increases slightly). The suppression of the total
excitation gap arises from the suppression of the order parameter. The
density of incoherent pairs, as measured by $n_p$, slightly increases
for not-so-large $n_i$. This supports the notion that nonmagnetic
impurities do not mainly break incoherent pairs. Instead, they scatter
the Cooper pairs out of the condensate.\cite{Note}

Our self-consistent calculations show that in the unitary limit, the low
$T$ superfluid density is quadratic in $T$ in a BCS $d$-wave
superconductor, in agreement with existing calculations and experiment.
Strong pair excitations add an additional $T^{3/2}$ term, with
preliminary experimental support. As a function of increasing $n_i$, the
zero $T$ superfluid density decreases faster at first for unitary
scatterers, whereas the opposite holds for scattering in the Born limit.
The former behavior is in agreement with experiment.

\acknowledgments

We would like to thank A.~V. Balatsky, P.~J. Hirschfeld, and K. Levin
for useful discussions. The numerics was in part carried out on the
computing facilities of the Department of Engineering, the Florida
State University. This work is supported by the State of Florida.

\appendix

\section{Impurity dressing for BCS theory at the Abrikosov-Gor'kov level}
\label{App:BCS}

\begin{figure}
\centerline{\includegraphics[width=3.4in]{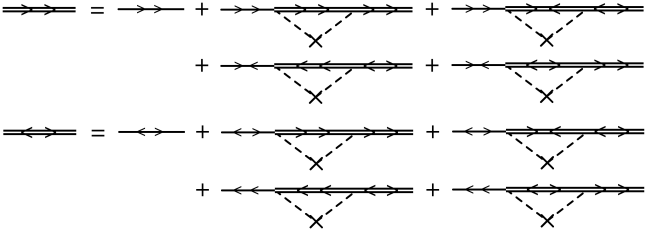}}
\caption{Impurity dressing at the Abrikosov-Gor'kov level in BCS theory.}
\label{diag:BCS-AG}
\end{figure}

In this appendix, we present the impurity dressing for a BCS
superconductor, following Abrikosov-Gor'kov\cite{AG,AGD}, but in a more
general form, namely, we do not assume $\bar{G}_{-\omega} =
-\bar{G}_\omega$. This will make it easier to understand the current
theory in the presence of strong pairing correlations, as there is a
strong similarity between the impurity dressing diagrams for both BCS
theory and the pairing fluctuation theory.

For a pure BCS superconductor, we have the Gor'kov equations,
\begin{subequations}
\begin{equation}
{G^0_0}^{-1}(K) G^0(K) = 1 - \Delta_{\bf k} {F^0}^\dag (K)\;, 
\end{equation}
\begin{equation}
{G^0_0}^{-1}(-K) {F^0}^\dag (K) =  \Delta^*_{\bf k} G^0 (K)\;.
\end{equation}
\label{eq:Gorkov-eq}
\end{subequations}
At the AG level, the relationship between the impurity dressed Green's
functions $G$ and $F$ is represented by the Feynman diagrams shown in
Fig.~\ref{diag:BCS-AG}. Define the impurity average $\bar{G}_\omega$ as
in Eq.~(\ref{eq:Gwbar}), and
\begin{equation}
\bar{F}^\dag_\omega = n_i \sum_{\bf k^\prime} |u(\textbf{k}
-\textbf{k}^\prime)|^2 F^\dag (K^\prime) \;, 
\label{eq:Fwbar}
\end{equation}
as well as their complex conjugate. Note Fig.~\ref{diag:BCS-AG} is
actually Fig.~105 in Ref.~\onlinecite{AGD}. Without giving details, we
give the result following AG:
\begin{subequations}
\begin{equation}
(i\omega -\ek -\bar{G}_\omega) G(K) + (\Delta_{\bf k} + \bar{F}_\omega)
F^\dag (K)  = 1\,,
\end{equation}
\begin{equation}
(i\omega + \ek + \bar{G}_{-\omega}) F^\dag (K) + (\Delta^*_{\bf k} +
\bar{F}^\dag_\omega)  G (K)  = 0\,.
\end{equation}
\label{eq:Gorkov-eq-imp}
\end{subequations}
Define $i\tilde\omega = i\omega - \bar{G}_\omega$,
$i\tilde{\underline{\omega}} = -i\omega - \bar{G}_{-\omega}$,
$\tilde{\Delta}_{\bf k} = \Delta_{\bf k} + \bar{F}_\omega$, and
$\tilde{\Delta}^*_{\bf k} = \Delta_{\bf k}^* +
\bar{F}^\dag_\omega$. Then we obtain
\begin{subequations}
\begin{equation}
G(K) = \frac{i\tilde{\underline{\omega}} -\ek}
{(i\tilde{\omega}-\ek) (i\tilde{\underline{\omega}} -\ek)  +
  \tilde{\Delta}_\mb{k}^* \tilde{\Delta}_\mb{k}} \,,
\end{equation}
\begin{equation}
F^\dag(K) = \frac{\tilde{\Delta}^*_\mb{k}}
{(i\tilde{\omega}-\ek) (i\tilde{\underline{\omega}} -\ek)  +
  \tilde{\Delta}_\mb{k}^* \tilde{\Delta}_\mb{k}} \,.
\end{equation}
\label{eq:BCS-AG}
\end{subequations}

For $d$-wave, the first equation becomes Eq.~(\ref{eq:Gfinal}). Note
$G(K)$ is no longer symmetrical in $\omega$ in general as a consequence
of impurity scattering, but $F(K)$ still is, since $F(K)$ involves $\pm
\omega$ pairs.

\begin{figure}
\centerline{\includegraphics[width=3in]{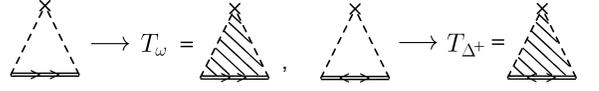}}
\caption{Replacement scheme from the AG level impurity scattering to 
self-consistent  impurity treatment in BCS theory.}
\label{diag:BCS-Replace}
\end{figure}

\begin{figure}[tbh]
\vspace*{2.ex}
\centerline{\includegraphics[width=3.2in]{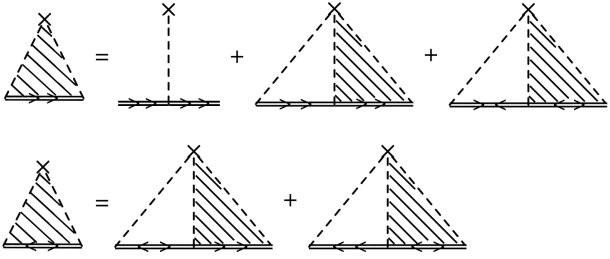}}
\caption{Relationship between impurity $T$ matrices $T_\omega$ and
  $T_{\Delta^\dag}$ in BCS theory.}
\label{diag:BCS-Tmatrix}
\end{figure}

The above result can be easily extended to self-consistent impurity
$T$-matrix calculations, by replacing the AG-level impurity scattering
with the self-consistent impurity $T$-matrices, as shown in
Fig.~\ref{diag:BCS-Replace}.  The relationship between the regular and
anomalous impurity $T$ matrices $T_\omega$ and $T_{\Delta^\dag}$ are shown
in Fig.~\ref{diag:BCS-Tmatrix}. One can easily write down the
corresponding equations, as follows.

\begin{subequations}
\begin{equation}
T_\omega = u + u\dbar{G}_\omega T_\omega - u \dbar{F}_\omega
T_{\Delta^\dag} \,,
\end{equation}
\begin{equation}
 T_{\Delta^\dag} = u \dbar{F}^\dag_\omega T_\omega + u
 \dbar{G}_{-\omega} T_{\Delta^\dag} \,.
\end{equation}
\label{eq:BCS-Tmatrix}
\end{subequations}
Finally, one has
\begin{subequations}
\label{eq:BCS-TSolutions}
\begin{equation}
T_\omega = \frac{u (1-u\dbar{G}_{-\omega})}{\left( 1-u
    \dbar{G}_\omega\right) \left( 1-u
    \dbar{G}_{-\omega}\right) + u^2 \dbar{F}_\omega
    \dbar{F}^\dagger_\omega} \,, 
\label{eq:BCS-Tw}
\end{equation}
\begin{equation}
T_{\Delta^\dag}(\omega) = \frac{u^2 \dbar{F}^\dagger_\omega} {\left( 1-u
    \dbar{G}_\omega\right) \left( 1-u
    \dbar{G}_{-\omega}\right) + u^2 \dbar{F}_\omega
    \dbar{F}^\dagger_\omega } \,,
\label{eq:BCS-TDelta}
\end{equation}
\end{subequations}
where $\dbar{G}=\sum_\mb{k} G(K)$, $\dbar{F}^\dag = \sum_\mb{k}
F^\dag(K)$, and similarly for their complex conjugate. Note these two
equations are formally identical to Eqs.~(\ref{eq:Tw}) and
(\ref{eq:TDelta}), except that the current $T_{\Delta^\dag}$ contains
the factor $\Delta$ already.

Now with the new definition $i\tilde\omega = i\omega - \Sigma_\omega$,
$i\tilde{\underline{\omega}} = -i\omega - \Sigma_{-\omega}$,
$\tilde{\Delta}_{\bf k} = \Delta_{\bf k} + \Sigma_\Delta$, and
$\tilde{\Delta}^*_{\bf k} = \Delta_{\bf k}^* +
\Sigma_\Delta^*$, as well as $\Sigma_\omega = n_i T_\omega$ and
$\Sigma^*_\Delta= n_i T_{\Delta^\dag}$,  
Eqs.~(\ref{eq:BCS-AG}) for $G$ and $F$ remain valid.


\begin{thebibliography}{43}
\bibitem[{Not()}]{Note}
\bibinfo{note}{Of course, when $n_i$ is too large, all pairs will be
  destroyed.}
\end{thebibliography}

\end{document}